# EXACT SOLUTIONS FOR THE RELATIVISTIC DYNAMICS OF A SELF-CONSISTENT SYSTEM WITH ELECTROMAGNETIC AND GRAVITATIONAL INTERACTION WITHIN THE WIGNER-VLASOV FORMALISM

E.E. Perepelkin[a,c,d], B.I. Sadovnikov[a], N.G. Inozemtseva[b,c], I.Yu. Baibara[a]

[a] *Faculty of Physics, Lomonosov Moscow State University, Moscow, 119991 Russia*
[b] *Moscow Technical University of Communications and Informatics, Moscow, 123423 Russia*
[c] *Dubna State University, Moscow region, Dubna,141980 Russia*
[d] *Joint Institute for Nuclear Research, Moscow region, Dubna,141980 Russia*

**Abstract**

This work derives exact solutions to the problem of interacting particle density evolution in relativistic and quasi-relativistic approximations for electromagnetic and gravitational interactions. Two types of radial symmetry for the initial density distribution are considered: spherical and cylindrical. It is shown that the relativistic effect delays the onset of the shock wave moment, and in some cases removes it entirely or, conversely, can facilitate it.

The analysis of the system's dynamics is carried out within the Wigner-Vlasov formalism, which makes it possible to extend the obtained solutions to quantum systems, including those with gravitational interaction. The derived exact solutions can be directly used as a cross-check for modeling and optimizing nonlinear problems of beam dynamics with account for space charge, astrophysics, plasma physics, and quantum systems with a shock wave.

## Introduction

In the mid-20th century, A.A. Vlasov constructed an infinite self-consistent chain of kinetic equations for the distribution functions $f_1(\vec{r},t)$, $f_2(\vec{r},\vec{v},t)$, $f_3(\vec{r},\vec{v},\dot{\vec{v}},t)$,… of kinematic quantities of all orders $\vec{r},\vec{v},\dot{\vec{v}},\ddot{\vec{v}},...$ [1-2]. The foundation of the Vlasov chain of equations is a single first principle – the law of probability conservation for the function $f_\infty(\vec{r},\vec{v},\dot{\vec{v}},...,t)$ in generalized phase space $\Omega_\infty = \{\vec{r},\vec{v},\dot{\vec{v}},\ddot{\vec{v}},...\}$ [3]. The first two equations of the Vlasov chain have gained the most prominence in hydro- and gas dynamics [4-6], plasma physics [7-9], solid-state physics [10-12], astrophysics [13-18], statistical physics [19-21], and accelerator physics [22-24].

$$\frac{\partial}{\partial t}f_1(\vec{r},t) + \mathrm{div}_r\left[f_1(\vec{r},t)\langle\vec{v}\rangle(\vec{r},t)\right] = 0, \qquad (\mathrm{i}.1)$$

$$\frac{\partial}{\partial t}f_2(\vec{r},\vec{v},t) + \mathrm{div}_r\left[\vec{v}f_2(\vec{r},\vec{v},t)\right] + \mathrm{div}_v\left[f_2(\vec{r},\vec{v},t)\langle\dot{\vec{v}}\rangle(\vec{r},\vec{v},t)\right] = 0, \qquad (\mathrm{i}.2)$$

where

$$f_1(\vec{r},t) = \int_{\mathbb{R}^3} f_2(\vec{r},\vec{v},t)d^3v, \quad f_2(\vec{r},\vec{v},t) = \int_{\mathbb{R}^3} f_3(\vec{r},\vec{v},\dot{\vec{v}},t)d^3\dot{v}, \qquad (\mathrm{i}.3)$$

$$f_1(\vec{r},t)\langle\vec{v}\rangle(\vec{r},t) = \int_{\mathbb{R}^3} \vec{v}f_2(\vec{r},\vec{v},t)d^3v, \quad f_2(\vec{r},\vec{v},t)\langle\dot{\vec{v}}\rangle(\vec{r},\vec{v},t) = \int_{\mathbb{R}^3} \dot{\vec{v}}f_3(\vec{r},\vec{v},\dot{\vec{v}},t)d^3\dot{v},$$

$$f_1(\vec{r},t)\langle\langle\dot{\vec{v}}\rangle\rangle(\vec{r},t) = \int_{\mathbb{R}^3}\langle\dot{\vec{v}}\rangle(\vec{r},\vec{v},t)f_2(\vec{r},\vec{v},t)d^3v.$$

Note that the fulfillment of relations (i.3) implies a sufficiently rapid decay of the distribution functions at infinity, which guarantees the convergence of the integrals.

The equations of the Vlasov chain contain conservation laws [25]. For instance, integrating the second equation (i.2) over velocity space yields the first equation (i.1), known as the continuity equation, which describes the law of mass or charge conservation. Multiplying equation (i.2) by a velocity component $v_k$ and subsequent integration over velocity space leads to the law of momentum conservation:

$$\frac{d}{dt}\langle v_k \rangle = \left( \frac{\partial}{\partial t} + \langle v_s \rangle \frac{\partial}{\partial x_s} \right) \langle v_k \rangle = -\frac{1}{f_1} \frac{\partial P_{ks}}{\partial x_s} + \langle\langle \dot{v}_k \rangle\rangle, \qquad (i.4)$$

$$P_{ks} = \int_{\mathbb{R}^3} (v_k - \langle v_k \rangle)(v_s - \langle v_s \rangle) f_2 d^3v, \qquad (i.5)$$

where $P_{ks}$ corresponds to the pressure tensor, and $\partial P_{ks}/\partial x_s$ represents the pressure force. The term $\langle\langle \dot{v}_k \rangle\rangle$ is associated with the external force acting on the system. Multiplying the second equation (i.2) by $v^2$ and subsequent integration over velocity space leads to the law of energy conservation:

$$\frac{\partial}{\partial t}\left[ \frac{f_1}{2}|\langle \vec{v} \rangle|^2 + \frac{1}{2}\text{Tr}\,P_{kk} \right] + \frac{\partial}{\partial x_s}\left[ \frac{f_1}{2}|\langle \vec{v} \rangle|^2 \langle v_s \rangle + \frac{1}{2}\langle v_s \rangle \text{Tr}\,P_{kk} + \langle v_k \rangle P_{ks} + \frac{1}{2}\text{Tr}\,P_{kks} \right] = \int_{\mathbb{R}^3} f_2 \langle \dot{v}_k \rangle v_k d^3v,$$

$$P_{kns} = \int_{\mathbb{R}^3} (v_k - \langle v_k \rangle)(v_n - \langle v_n \rangle)(v_s - \langle v_s \rangle) f_2 d^3v, \qquad (i.6)$$

where summation over repeated indices is implied. The first term in equation (i.6) accounts for the change in energy density over time. The second term in (i.6) corresponds to the divergence of the energy flux density, and the right-hand side is associated with the power of external forces. Equations (i.4)-(i.6) are written in the Cartesian coordinate system, hence covariant and contravariant indices are equivalent.

The Vlasov chain of equations is self-linking, meaning that to solve equation (i.1), one needs to know the average velocity field which, according to (i.3), is determined by the function $f_2$, satisfying the second Vlasov equation (i.2). To solve equation (i.2), one needs to know the average acceleration field $\langle \dot{\vec{v}} \rangle$, which, according to (i.3), is determined by the function $f_3$, satisfying the third Vlasov equation. Thus, solving the $n$-th equation requires solving the $n+1$-th equation. In practice, this situation is resolved by cutting off the chain of equations via the introduction of a dynamic approximation for the average kinematic quantities $\langle \vec{v} \rangle$, $\langle \dot{\vec{v}} \rangle$ and so on.

When cutting the Vlasov chain off at the first equation (i.1), one can utilize the Helmholtz decomposition of the vector field $\langle \vec{v} \rangle$ into potential $\nabla_r \Phi$ and vortex $\vec{A}_\Psi$ field [26, 27]

$$\langle \vec{v} \rangle(\vec{r},t) = -\alpha_1 \nabla_r \Phi(\vec{r},t) + \alpha_2 \vec{A}_\Psi(\vec{r},t), \qquad (i.7)$$

where $\alpha_1, \alpha_2 \in \mathbb{R}$ are some constant quantities. Using the positivity of the function $f_1 \geq 0$, we represent it as $f_1 = |\Psi|^2$, where $\Psi \in \mathbb{C}$. The complex function $\Psi = |\Psi|\exp(i\varphi)$ has two parameters $|\Psi|$ and the phase $\varphi$. The modulus $|\Psi|$ is related to $f_1$, and the phase $\varphi$ is a free

parameter. Since both parameters of the function $\Psi$ are related to the first equation, we define the phase $\varphi$ via a scalar potential $\Phi$ (i.7)

$$\Phi(\vec{r},t) \stackrel{\text{def}}{=} 2\varphi(\vec{r},t) + 2\pi k, \, k \in \mathbb{Z}, \qquad \langle \vec{v} \rangle(\vec{r},t) = i\alpha_1 \nabla_r \operatorname{Ln}\left(\frac{\Psi}{\Psi^*}\right) + \alpha_2 \vec{A}_\Psi, \tag{i.8}$$

where $\operatorname{Ln} z \stackrel{\text{def}}{=} \ln|z| + i \operatorname{Arg} z$, $\operatorname{Arg} z = \arg z + 2\pi k$, $k \in \mathbb{Z}$. The direct substitution of the expression (i.8) for the average velocity $\langle \vec{v} \rangle$ and the representation $f_1 = |\Psi|^2$ into equation (i.1) leads to a $\Psi$-Schrödinger-type equation

$$\frac{i}{\alpha_3} \frac{\partial \Psi}{\partial t} = -\alpha_1 \alpha_3 \left( \hat{p} - \frac{\alpha_2}{2\alpha_1 \alpha_3} \vec{A}_\Psi \right)^2 \Psi + U\Psi, \tag{i.9}$$

where the operator $\hat{p} \stackrel{\text{def}}{=} -\frac{i}{\alpha_3} \nabla_r$, $\alpha_3 \neq 0$, $\alpha_3 \in \mathbb{R}$ is a constant; $U(\vec{r},t) \in \mathbb{R}$ is a certain function. Substituting $\Psi = |\Psi| \exp(i\varphi)$ into equation (i.9) yields a $\Psi$ - Hamilton-Jacobi-type equation

$$-\frac{1}{\alpha_3} \frac{\partial \varphi}{\partial t} = -\frac{1}{4\alpha_1 \alpha_3} |\langle \vec{v} \rangle|^2 + V \stackrel{\text{def}}{=} H, \qquad V = U + Q, \qquad Q \stackrel{\text{def}}{=} \frac{\alpha_1}{\alpha_3} \frac{\Delta_r |\Psi|}{|\Psi|}, \tag{i.10}$$

in which the average velocity field $\langle \vec{v} \rangle$ satisfies the equation of motion

$$\frac{d}{dt} \langle \vec{v} \rangle = -\alpha_2 \left( \vec{E}_\Psi + \langle \vec{v} \rangle \times \vec{B}_\Psi \right), \qquad \vec{E}_\Psi \stackrel{\text{def}}{=} -\frac{\partial \vec{A}_\Psi}{\partial t} - \frac{2\alpha_1 \alpha_3}{\alpha_2} \nabla_r V, \qquad \vec{B}_\Psi \stackrel{\text{def}}{=} \operatorname{curl}_r \vec{A}_\Psi. \tag{i.11}$$

Furthermore, for the vortex field $\vec{A}_\Psi$ and the potential V one can write down Lorenz $\Psi$-gauge:

$$\operatorname{div}_r \vec{A}_\Psi + \frac{2\alpha_1 \alpha_3}{\alpha_2} \frac{1}{c^2} \frac{\partial V}{\partial t} = 0. \tag{i.12}$$

Note that expressions (i.9)-(i.12) are a special case of the first Vlasov equation (i.1) with the Helmholtz decomposition (i.8). The following values can be taken as the arbitrary constants $\alpha_j$, $j = 1...3$:

$$\alpha_1 = -\frac{\hbar}{2m} = -\frac{c \lambdabar_c}{2}, \qquad \alpha_2 = -\frac{q}{m}, \qquad \alpha_3 = \frac{1}{\hbar}. \tag{i.13}$$

where $\hbar$ is the reduced Planck constant, $\lambdabar_c$ is the reduced Compton wavelength; $q, m$ are the charge and mass of the particle, respectively. In the absence of an electric charge $q$, one can take an analog of the gravitational charge $q_g = m\sqrt{4\pi\varepsilon_0 G}$, then substituting the values from (i.13) into equation (i.9) transforms it into the Schrödinger equation for a particle in an electromagnetic field $\vec{E}_\Psi, \vec{B}_\Psi$ with scalar potential $\varphi_\Psi = V/q$ and vector potential $\vec{A}_\Psi$. Equation (i.10) transforms into the well-known Hamilton-Jacobi equation. The function H is the Hamiltonian, and the function Q is the quantum potential used in the de Broglie–Bohm «pilot-wave» theory [28-30]. Equation (i.11) corresponds to the equation of motion of a charged particle in an electromagnetic field under the action of the Coulomb force and the Lorentz force. Since the potential V onsists of a superposition of two potentials (i.11) $U$ and $Q$ — the gauge condition (i.12) can be represented as the sum of two gauge conditions

$$\text{div}_r \vec{A} + \frac{1}{c^2}\frac{\partial \varphi}{\partial t} = 0, \qquad \text{div}_r \vec{A}_Q + \frac{1}{qc^2}\frac{\partial Q}{\partial t} = 0, \qquad (i.14)$$

where

$$\vec{A}_\Psi \stackrel{\text{def}}{=} \vec{A} + \vec{A}_Q, \quad \vec{B}_\Psi = \text{curl}_r \vec{A} + \text{curl}_r \vec{A}_Q = \vec{B} + \vec{B}_Q, \qquad (i.15)$$

$$\vec{E}_\Psi = -\frac{\partial \vec{A}}{\partial t} - \frac{1}{q}\nabla_r U - \frac{\partial \vec{A}_Q}{\partial t} - \frac{1}{q}\nabla_r Q = \vec{E} + \vec{E}_Q.$$

The gauge for the potentials $\varphi$ and $\vec{A}$ is known as the Lorenz gauge. The gauge for the quantum potential $Q = q\varphi_Q$ and the vector potential $\vec{A}_Q$ has a similar form but is related to the quantum statistical nature of the system. Indeed, in the classical limit as $\hbar \ll 1$ the coefficient $\alpha_1/\alpha_3 = \hbar^2/2m$ in the quantum potentia Q becomes small (i.10), (i.e. $Q \approx 0$). Consequently, only deterministic electromagnetic fields $\vec{E}$ and $\vec{B}$ are present in the potential in the Hamilton-Jacobi equation (i.10) $V \approx U$ and in the equation of motion (i.11). The fields $\vec{E}_Q$ and $\vec{B}_Q$ are associated with the quantum nature of the system.

Note that the fields $\vec{E}_\Psi, \vec{B}_\Psi$ generally do not satisfy Maxwell's equations. On the one hand, two of Maxwell's equations follow from relations (i.11):

$$\text{div}_r \vec{B}_\Psi = 0, \quad \text{curl}_r \vec{E}_\Psi = -\frac{\partial \vec{B}_\Psi}{\partial t}. \qquad (i.16)$$

The remaining two Maxwell's equations can be obtained if the condition of self-consistency for the system is fulfilled

$$\rho_\Psi \stackrel{\text{def}}{=} \varepsilon_0 \text{div}_r \vec{E}_\Psi \stackrel{\text{def}}{=} \text{div}_r \vec{D}_\Psi = qf_1. \qquad (i.17)$$

This condition is essentially one of Maxwell's equations. Taking into account the first Vlasov equation (i.1), the last equation follows from it

$$\frac{\partial}{\partial t}\vec{D}_\Psi + \vec{J}_\Psi = \text{curl}_r \vec{H}_\Psi, \qquad (i.18)$$

where $\vec{J}_\Psi \stackrel{\text{def}}{=} \rho_\Psi \langle \vec{v} \rangle$, $\vec{H}_\Psi$ is a certain field, for example $\mu_0 \vec{H}_\Psi \stackrel{\text{def}}{=} \vec{B}_\Psi$. Thus, provided the self-consistency condition (i.17) is satisfied, the system of Maxwell's equations with the gauge (i.12) holds.

$$\begin{cases} \Box \varphi_\Psi = \dfrac{\rho_\Psi}{\varepsilon_0}, \\ \Box \vec{A}_\Psi = \mu_0 \vec{J}_\Psi, \end{cases} \Rightarrow \Box A_\Psi^\nu = \mu_0 J_\Psi^\nu, \qquad (i.19)$$

$$\varphi_\Psi = \varphi + \varphi_Q, \quad \vec{J}_\Psi = \vec{J} + \vec{J}_Q, \quad \rho_\Psi = \rho + \rho_Q,$$

where $\Box = \partial^\mu \partial_\mu$ is the d'Alembert operator with the metric $g_{\mu\nu} \mapsto (+,-,-,-)$, $A_\Psi^\nu = \left(\varphi_\Psi/c, \vec{A}_\Psi\right)$, $J_\Psi^\nu = \left(c\rho_\Psi, \vec{J}_\Psi\right)$ are the 4-potential and 4-current vectors, respectively.

From a physical point of view, the self-consistency condition (i.17) requires accounting for not only the influence of external fields on the system but also the influence of the system itself on the external fields. Thus, the external fields and the system itself form a single whole with mutual influence on each other. It should be noted that from a practical standpoint, finding an exact solution to a self-consistent problem in the sense of fulfilling condition (i.17) is very rarely achievable.

For example, a classical problem for the Schrödinger equation (i.9) is to find the wave function $\Psi$ or a given (fixed) external potential $U$. The influence of the quantum system itself on the potential $U$ is considered negligible.

A similar problem arises in electrodynamics described by Maxwell's equations (i.19). Finding the electromagnetic fields reduces to initially specifying the 4-current $J_\Psi^\nu$, from which the 4-potential $A_\Psi^\nu$ is found. In the case of a self-consistent problem, the current sources $J_\Psi^\nu$ themselves depend on the fields created by them. In practice, to resolve this problem, for instance, in accelerator physics when modeling the dynamics of a charged particle beam taking into account the space charge effect [31-33], Maxwell's equations are supplemented with an equation of motion of the type (i.11). As a result, the solution reduces to a numerical iterative scheme, in which, for fixed currents $J_\Psi^\nu$, the fields are first found, and then the currents $J_\Psi^\nu$ are recalculated for the next time moment using the equations of motion.

This gives rise to a number of problems related to the equation of motion itself. In the non-relativistic case, one can consider equation (i.11), derived from the first principle, but within the non-relativistic Helmholtz decomposition (i.7). Despite the fact that the invariance of Maxwell's equations (i.19) incorporates Lorentz transformations and, essentially, special relativity, they do not allow one to solve the problem of electrodynamics without an equation of motion. A formal rewriting of the equation of motion (i.11) with the relativistic momentum on the left-hand side, $\langle \vec{p} \rangle = m\gamma \langle \vec{v} \rangle$ ($\gamma^{-2} = 1 - \beta^2$) does not solve the problem, not least because it does not include the effect of electromagnetic radiation. As is known, a charged particle moving with acceleration emits an electromagnetic wave. The presence of electromagnetic radiation leads to a radiation reaction force proportional to $\langle \dddot{v} \rangle$, which results in a third-order differential equation of motion, for example, the Lorentz-Abraham-Dirac equation [34-35]. Note that such equations of motion, unlike (i.11), are phenomenological and contain a number of peculiar unphysical solutions requiring careful consideration.

The equations of motion (i.4) and (i.11) are derived from the first principle based on the first and second Vlasov equations. Consequently, their right-hand sides must be equal. If magnetic fields are absent, the following relation holds:

$$\frac{1}{m}\frac{\partial Q}{\partial x^k} = \frac{1}{f_1}\frac{\partial P_{ks}}{\partial x^s}, \quad \langle\langle \dot{v}_k \rangle\rangle = -\frac{1}{m}\frac{\partial U}{\partial x^k}. \qquad (i.20)$$

From expressions (i.20), it follows that the quantum potential $Q$ creates a quantum pressure that counteracts the external force determined by the quantity $\langle\langle \dot{v}_k \rangle\rangle$. For example, when considering a quantum system with a potential well $U$, the external potential $U$ will «strive» to localize the system inside the well with a force $-\nabla_r U$, while the quantum pressure will oppose it with a force $-\nabla_r Q$. As a result, the system will be in «equilibrium» with a zero average velocity $\langle \vec{v} \rangle = \vec{0}$.

Around 1932, in the works of Weyl and Wigner, the quasi-probability density function for quantum systems in phase space was introduced [36-37]:

$$W(\vec{r},\vec{p},t) = \frac{1}{(2\pi\hbar)^3} \int_{\mathbb{R}^3} \Psi^*\left(\vec{r}-\frac{\vec{s}}{2},t\right) \Psi\left(\vec{r}+\frac{\vec{s}}{2},t\right) e^{-\frac{i}{\hbar}\vec{s}\cdot\vec{p}} d^3s. \qquad (i.21)$$

If the system lacks a magnetic field, then the Wigner function (i.21) satisfies the Moyal evolution equation [38]:

$$\frac{\partial W}{\partial t} + \frac{1}{m}\vec{p}\cdot\nabla_r W - \nabla_r U \cdot \nabla_p W = \sum_{l=1}^{+\infty} \frac{(-1)^l (\hbar/2)^{2l}}{(2l+1)!} U\left(\vec{\nabla}_r \cdot \vec{\nabla}_p\right)^{2l+1} W, \qquad (i.22)$$

which is a special case of the second Vlasov equation (i.2) with the Vlasov-Moyal approximation for the average vector acceleration field [39]:

$$f_2 \langle \dot{v}_k \rangle = \sum_{l=0}^{+\infty} \frac{(-1)^{l+1}(\hbar/2)^{2l}}{m^{2l+1}(2l+1)!} \frac{\partial U}{\partial x^k} \left(\vec{\nabla}_r \cdot \vec{\nabla}_v\right)^{2l} f_2 \;\Rightarrow\; \langle\langle \dot{v}_k \rangle\rangle = -\frac{1}{m}\frac{\partial U}{\partial x^k}, \qquad (i.23)$$

where $f_2(\vec{r},\vec{v},t) = m^3 W(\vec{r},\vec{p},t)$. The arrows over the operators in expressions (i.22)-(i.23) indicate the direction of their action. A peculiarity of the Wigner function (i.21) is the presence of negative values for wave functions different from a Gaussian distribution (Hudson's theorem [40] and its extensions to the 3D case [41]). The presence of negative values of the Wigner function does not contradict the second Vlasov equation, as no positivity condition was imposed on the distribution functions during the construction of the Vlasov chain. In the presence of a magnetic field, an extended Vlasov-Moyal approximation [42] or the gauge-invariant Weyl-Stratonovich function [43-47] can be used.

A comparison of the Vlasov-Moyal approximation (i.23) and expression (i.20) reveals their correspondence. Indeed, averaging (i.23) over velocity space yields expression (i.20). Also, in the classical limit $\hbar \ll 1$ only the first term $-\frac{1}{m}\frac{\partial U}{\partial x^k}$ remains in the approximation (i.23). Thus, the Vlasov-Moyal approximation (i.23), obtained via the Wigner function (i.21), has a direct connection with expressions (i.20) derived from the Vlasov equations.

Note that the form of the approximation (i.23) is not unique [42]. There exists an infinite number of ways to construct the sum of the series (i.23). Nevertheless, all of them, when averaged over velocity space, will lead to the same expression $\langle\langle \dot{v}_k \rangle\rangle$ (i.23). A similar result will hold if one performs the classical limit $\hbar \to 0$ provided the derivatives in the series (i.23) are bounded. This ambiguity is a fundamental property of the Vlasov equations, which contain only the operators $\mathrm{div}_r$, and do not contain the operators $\mathrm{curl}_r$. As a result, by the Helmholtz decomposition theorem, it is impossible to uniquely reconstruct the vector field $\langle \dot{v}_k \rangle$. This situation is analogous to the Heisenberg uncertainty principle: on a macro-scale, the concept of a trajectory defined by the equation of motion (i.4) with a deterministic, unique external force $m\langle\langle\dot{v}_k\rangle\rangle$ whereas on a micro-scale, the «force» $m\langle\dot{v}_k\rangle$ is not uniquely definable due to the arbitrariness in the value of the solenoidal field.

The distribution functions $f_n$ included in the Vlasov chain of equations can have different physical interpretations depending on the problem being solved. For example, in continuum mechanics, the function $f_1$ can be interpreted as the mass density function $\rho = mf_1$, while in electrodynamics (plasma physics) one can consider the charge density $\rho = qf_1$. From the standpoint of quantum mechanics, the function $f_1$ can be understood as the probability density $f_1 = |\Psi|^2$, and the function $f_2$ will correspond to the Wigner function. The second Vlasov equation (i.2) with the approximation (i.23) in the classical limit $\hbar \ll 1$ transforms into the classical Liouville equation. The equations described above naturally (from a single first principle – the law of probability conservation) connect various branches of classical and quantum physics.

The aim of this work is to construct exact self-consistent solutions for problems in electrodynamics and gravity within the framework of the presented Wigner-Vlasov formalism in the classical and relativistic approximations, as well as their interpretation from the standpoint of quantum mechanics.

The work is structured as follows. §1 considers the construction of an exact analytical solution for the self-consistent space charge problem. For systems with spherically (p.1.1) and cylindrically (p.1.2) symmetric initial charge density distributions, considering relativism, characteristic equations describing the motion of non-intersecting spherical and cylindrical layers are derived. The charge density function evolutions are constructed based on the characteristic equations. The system's dynamics in the relativistic and classical limits are described in detail. The presence of inhomogeneity in the initial distribution leads to solutions in the form of a shock wave (Coulomb explosion) [48-53]. It is shown that accounting for relativism causes a delay in the shock wave formation time. For homogeneous initial distributions, the picture of density evolution changes fundamentally when relativistic effects are taken into account. §2 considers a self-consistent problem in the quasi-relativistic approximation concerning the gravitational collapse of systems with spherically (p.2.1) and cylindrically (p.2.2) symmetric initial mass density distributions. In the case of an inhomogeneous initial distribution, the presence of a shock wave solution leading to the formation of a high-density spherical shell is demonstrated. For a homogeneous initial distribution, accounting for quasi-relativism leads to the absence of a shock wave in the center of the system and a fundamentally different form of the mass density distributions at large times compared to the non-relativistic limit.

§3 presents a discussion on describing the obtained exact solutions from §1-2 in terms of the Wigner-Vlasov formalism. Questions regarding the applicability of the formalism to relativistic systems and possible interpretations of their behavior are considered. The applicability of the $\Psi$ - Schrödinger equation for quasi-relativistic gravitational systems is discussed. The Conclusion contains the main results of the work. Appendices A and B contain proofs of the theorems.

**§1 The electromagnetic case**

Let us consider the evolution of the charge density distribution under the action of its own electric and magnetic fields. As an example, we take two types of symmetry for the initial distribution $\rho_0(\vec{r}) = \rho(\vec{r},0)$: spherical and cylindrical.

## 1.1 Spherically symmetric charge density distribution

Consider a system of concentric spheres centered at the origin of the coordinate system, covering the domain of the charge density distribution $\rho_s(r,t) = q f_1(r,t)$. For definiteness, we will assume that the function $\rho_s(r,t)$ is positive (or negative) throughout the coordinate space $\vec{r} \in \mathbb{R}^3$. At the initial moment of time, inside each such sphere of radius $R_0$, there is a charge $qN_s(R_0)$. Over time, due to the forces of electric repulsion, the radius of the sphere $R(t, R_0)$ will increase, but the charge contained within it will remain unchanged, i.e.,

$$qN_s(R_0) = 4\pi \int_0^{R_0} \rho_s(x,0) x^2 dx = 4\pi \int_0^{R(t,R_0)} \rho_s(x,t) x^2 dx = q\, \mathrm{N}_s\big[R(t,R_0),t\big]. \qquad (1.1)$$

**Remark 1** Note that expression (1.1) is valid provided the spheres do not intersect with each other over time. As will be shown below, depending on the initial density distribution $\rho_0(r)$, at some moment $t = t_c$ it is possible for two initially distinct $R_1, R_2$ spherical layers $R(t_c, R_1) = R(t_c, R_2)$ to intersect. In this case, the charge located between the two spheres ends up in an infinitesimally small volume, consequently, the density $\rho_s \to \infty$. This situation is known as the «Coulomb explosion». Thus, within the framework of this problem statement, we consider the evolution of $\rho_s(r,t)$ up to the moment $t < t_c$ of shock wave formation.

Due to the radial symmetry of the charge density $\rho_s(r,t)$ only the normal component of the electric field $\vec{D}(r,t)$ exists on the surface of the sphere $R(t, R_0)$, for which the Maxwell equation $\mathrm{div}_r\, \vec{D} = \rho_s$ holds. Note that $\oint_\Gamma \vec{E} \cdot d\vec{l} = 0$ for any closed contour $\Gamma$. Indeed, the part of the contour $d\vec{l}$ lying on the sphere's surface has $\vec{E} \perp d\vec{l}$, while the radial parts, $d\vec{l} \parallel \vec{e}_r$ due to the radial symmetry of $\vec{E}(r,t)$, mutually cancel each other when integrated in opposite directions. Consequently, according to the Maxwell equation $\mathrm{curl}_r\, \vec{E} = -\partial \vec{B}/\partial t = \vec{0}$, thus, due to the radial symmetry of the dynamical system $\vec{B}(\vec{r},t) = \vec{0}$. Therefore, for a test charge $q$ located on the surface of a sphere of radius $R(t, R_0)$, taking into account (1.1), the relativistic equation of motion is valid:

$$\frac{d}{dt} \frac{\dot{R}}{\sqrt{1 - \dot{R}^2/c^2}} = \frac{\vartheta_s^2}{R^2}, \quad \vartheta_s^2(R_0) \stackrel{\text{def}}{=} r_q c^2 N_s(R_0), \quad r_q \stackrel{\text{def}}{=} \frac{q^2}{4\pi\varepsilon_0 m c^2}, \qquad (1.2)$$

where $m$ is the rest mass of a particle with charge $q$, $\varepsilon_0$ is the vacuum permittivity, and $c$ is the speed of light. The quantity $r_q$ corresponds to the «classical radius» of a particle with charge $q$ and mass $m$. In the particular case of an electron, $r_q$ coincides with the classical electron radius $r_e$, i.e. $r_q = r_e$ [54-55]. Equation (1.2) must be supplemented with initial conditions:

$$R(0, R_0) = R_0, \quad \dot{R}(0, R_0) = \dot{R}_0. \qquad (1.3)$$

**Remark 2** The equation of motion (1.2) does not account for the effect of electromagnetic radiation. The radiation reaction force is proportional to $F_{rad} \sim \dot{\vec{v}} = \dddot{R}$, which leads to the Lorentz-Abraham-Dirac equation, which is of the third order. In this case, the initial conditions (1.3) must also be extended by adding a condition on the initial acceleration $\ddot{R}(0, R_0) = \ddot{R}_0$.

**Theorem 1** *The Cauchy problem (1.2)-(1.3) has a solution* $R(t, R_0)$

$$C_0 \sqrt{\left(R - \frac{\bar{\beta}_s^2}{C_0 - 1}\right)\left(R - \frac{\bar{\beta}_s^2}{C_0 + 1}\right)} + \frac{\bar{\beta}_s^2}{C_0^2 - 1} \operatorname{arcch} C_0 \left(\frac{C_0^2 - 1}{\bar{\beta}_s^2 C_0} R - 1\right) + C_1 = ct\sqrt{C_0^2 - 1}. \quad (1.4)$$

*where* $\bar{\beta}_s(R_0) = \vartheta_s(R_0)/c$ *and* $C_0, C_1$ *are constants determined by the initial conditions (1.3):*

$$C_0 = \eta_s^2(R_0) + \left(1 - \frac{\dot{R}_0^2}{c^2}\right)^{-1/2}, \quad \eta_s^2(R_0) \stackrel{\text{def}}{=} \frac{\bar{\beta}_s^2(R_0)}{R_0}, \quad (1.5)$$

$$\frac{C_1}{R_0} = -C_0 \sqrt{\left(1 - \frac{\eta_s^2(R_0)}{C_0 - 1}\right)\left(1 - \frac{\eta_s^2(R_0)}{C_0 + 1}\right)} - \frac{\eta_s^2(R_0)}{C_0^2 - 1} \operatorname{arcch} C_0 \left(\frac{C_0^2 - 1}{\eta_s^2(R_0) C_0} - 1\right). \quad (1.6)$$

The proof of Theorem 1 is given in Appendix A.

**Corollary 1** *In the special case of zero initial velocity* $\dot{R}_0 = 0$, *the solution (1.4) takes the form:*

$$\mathcal{F}_s\left[\frac{R(t, R_0)}{R_0}, \eta_s^2(R_0)\right] = \lambda_s(R_0) t, \quad (1.7)$$

*where*

$$C_0 = 1 + \eta_s^2(R_0), \quad C_1 = 0, \quad \lambda_s(R_0) \stackrel{\text{def}}{=} \frac{c\eta_s(R_0)\sqrt{2 + \eta_s^2(R_0)}}{R_0\left[1 + \eta_s^2(R_0)\right]}, \quad (1.8)$$

$$\mathcal{F}_s(x, y) \stackrel{\text{def}}{=} \sqrt{(x-1)\left(x - \frac{y}{y+2}\right)} + \frac{\operatorname{arcch}[(y+2)x - y - 1]}{(y+1)(y+2)}. \quad (1.9)$$

The proof of Corollary 1 is given in Appendix A.

Note that the solution (1.7) can be rewritten using the inverse function $\mathcal{P}_s = \mathcal{F}_s^{-1}$. In this case, the coordinate $R(t, R_0)$ of the spherical layer is expressed in terms of time $t$. From the representation (1.7), we obtain

$$R(t, R_0) = R_0 \mathcal{P}_s\left[\lambda_s(R_0) t, \eta_s^2(R_0)\right]. \quad (1.10)$$

**Corollary 2** *In the non-relativistic limit* $c \to +\infty$, *the solution (1.7) takes the form:*

$$\lim_{c \to +\infty} \mathcal{F}_s(x, \eta_s^2) = \sqrt{x(x-1)} + \operatorname{arcch} \sqrt{x} = F_s(x), \quad (1.11)$$

$$\frac{t\vartheta_s(R_0)\sqrt{2}}{R_0^{3/2}} = F_s\left[\frac{R(t, R_0)}{R_0}\right]. \quad (1.12)$$

$$\lim_{c \to +\infty} \mathcal{P}_s\left[\lambda_s(r)t, \eta_s^2(r)\right] = P_s\left[\lambda_s(r)t\right], \quad \lim_{c \to +\infty} \lambda_s(r) = \lambda_s(r) = \frac{\vartheta_s(r)\sqrt{2}}{r^{3/2}}, \qquad (1.13)$$

$$\lim_{c \to +\infty} R(t,r) = \mathrm{R}(t,r) = rP_s\left[\lambda_s(r)t\right],$$

*where the function $P_s$ is the inverse of the function $F_s$, i.e. $P_s\left[F_s(y)\right] = x$.*

The proof of Corollary 2 is given in Appendix A.

Figure 1 shows the graphs of the characteristics (1.7) for two types of initial charge density distributions

$$\rho_0 = const, \qquad (1.14)$$

$$\rho_0(r) = \frac{qf_0\tau}{4\pi r^2}\rho_n(\tau r), \quad \rho_n(r) = \frac{1}{\sqrt{2\pi}\sigma_r r}\exp\left\{-\frac{1}{2\sigma_r^2}\left[\ln(r) - \mu_r\right]^2\right\}, \qquad (1.15)$$

where $\tau$ is a free parameter, $qf_0$ is the total charge of the system, and $\mu_r$, $\sigma_r$ are the mean and standard deviation, respectively, for the log-normal distribution ($\tau = 2$) (1.15). Without loss of generality, the values $\mu_r = 0$ and $\sigma_r = 0.2$ were taken. For these values of $\mu_r$ and $\sigma_r$ the log-normal distribution (1.15) resembles a charged spherical shell with a radius of $\sim 0.3$ in relative units.

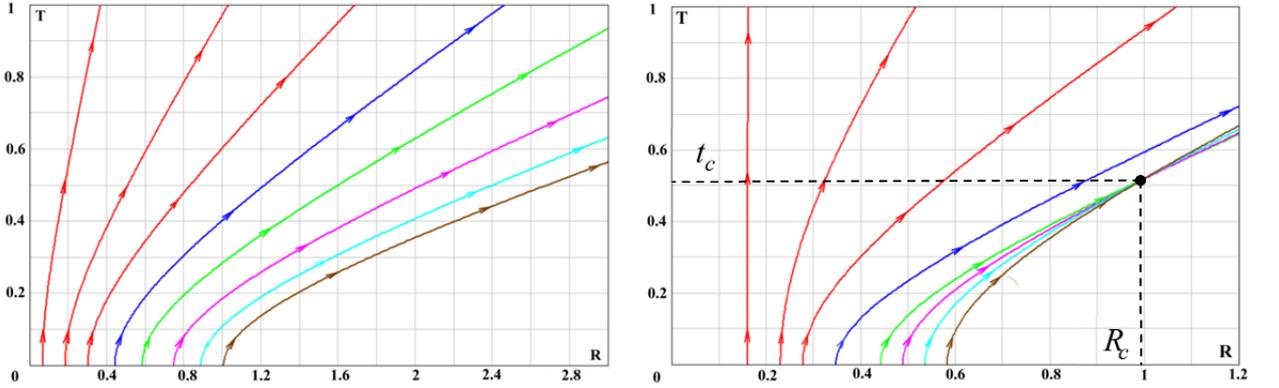

Fig. 1 Graphs of the characteristics (1.7) for distributions (1.14)-(1.15)

The left panel of Fig. 1 shows the characteristics for case (1.14), and the right panel of Fig. 1 shows the characteristics for case (1.15). Due to the Coulomb repulsion forces of like charges, the radius of the concentric spheres increase over time (see Fig. 1). The arrows in Fig. 1 indicate the direction of evolution of the radius of the spherical layers. Unlike the behavior of the characteristics in the left panel of Fig. 1, the characteristics for distribution (1.15) in the right panel have an intersection point $t = t_c$. The intersection of characteristics leads to zero volume occupied by the charge between the concentric layers, i.e., to an infinite increase in charge density.

Figure 2 shows a comparison of the graphs of characteristics (1.7) and (1.12) for the log-normal distribution (1.15). The red solid lines show the characteristics for the system with relativism (1.7), and the blue dashed lines show the characteristics for the classical system (1.12). Accounting for the relativistic effect leads to a slowdown in the expansion process of the charged system. The time of shock wave formation increases $t_c^{(relativistic)} > t_c^{(classic)}$. This effect has an intuitive classical interpretation in terms of relativistic mass $m_r(v_s) = m\gamma(v_s)$, where $\gamma(v_s) = (1-\beta_s^2)^{-1/2}$, $\beta_s = v_s/c$. The velocity magnitude $v$ in the considered problem statement (1.2) is $v_s = \dot{R}$. The dependence of the relativistic mass $m_r(v_s)$ on velocity leads to its «increase», thereby slowing down the motion of the concentric spheres under the same Coulomb force. As a result, the moment of shock wave formation in the relativistic case in Fig. 2 occurs later. When using the term relativistic mass $m$, one must keep in mind the remarks regarding its physical meaning [56].

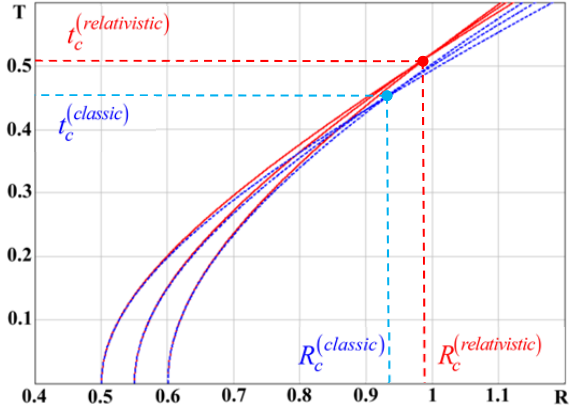

Fig. 2 The effect of relativism

**Corollary 3.** *From the representation (1.10) of the equation of motion, it follows that for a spherical layer with initial radius $R_0$ and zero initial velocity $\dot{R}_0 = 0$, the evolution $\beta_s(t, R_0) = v_s(t, R_0)/c$ and its time asymptotics are given by the expressions:*

$$\beta_s(t, R_0) = \eta_s(R_0) \frac{\sqrt{\kappa_s(t, R_0)\left[2 + \eta_s^2(R_0)\kappa_s(t, R_0)\right]}}{1 + \eta_s^2(R_0)\kappa_s(t, R_0)}, \qquad (1.16)$$

$$\beta_s^\infty(R_0) = \lim_{t \to +\infty} \beta_s(t, R_0) = \eta_s(R_0) \frac{\sqrt{2 + \eta_s^2(R_0)}}{1 + \eta_s^2(R_0)}, \qquad (1.17)$$

*where*

$$\kappa_s(t, R_0) = \frac{\mathcal{P}_s\left[\lambda_s(R_0)t, \eta_s^2(R_0)\right] - 1}{\mathcal{P}_s\left[\lambda_s(R_0)t, \eta_s^2(R_0)\right]}, \qquad (1.18)$$

*and in the non-relativistic case:*

$$v_s(t, R_0) = R_0 \lambda_s(R_0) \sqrt{\frac{P_s\left[\lambda_s(R_0)t\right] - 1}{P_s\left[\lambda_s(R_0)t\right]}}, . \qquad (1.19)$$

$$v_s^\infty(R_0) = \lim_{t \to +\infty} v_s(t, R_0) = R_0 \lambda_s(R_0). \qquad (1.20)$$

The proof of Corollary 3 is given in Appendix A.

**Remark 3** Note that the time asymptotics (1.17) and (1.20) have physical meaning only up to the time $t < t_c$. In the right panel of Fig. 1, the characteristics (1.10) intersect at $t = t_c$, so the limiting values (1.17) or (1.20) for such systems contain little physical information. On the other hand, for the homogeneous distribution (1.14), the characteristics (1.10) exhibit non-intersecting behavior at large times (see the left panel of Fig. 1). For such systems, the asymptotics (1.17) and (1.20) are informative. Indeed, it follows directly from (1.17) that none of the spherical layers will ever reach the speed of light ($\forall R_0 : \beta_s^\infty(R_0) \neq 1$), which is in full agreement with the theory of relativity, as the layer has non-zero mass. For each limiting value $0 \leq \beta_s^\infty < 1$ one can find the initial radius of the layer $R_0$ using the expression

$$\eta_s^2(R_0) = \frac{q^2 N_s(R_0)}{4\pi\varepsilon_0 mc^2 R_0} = \left[1-\left(\beta_s^\infty\right)^2\right]^{-1/2} - 1 = \gamma\left(v_s^\infty\right) - 1, \qquad (1.21)$$

which follows directly from the asymptotics (1.17). If the charge $qN_s(R_0)$ is approximated by a collection of $N_e$ electron charges ($q = q_e$, $m = m_e$), i.e., $qN_s(R_0) = N_e q_e$, then the left-hand side of expression (1.21) can be rewritten in terms of the classical electron radius $r_e = q_e^2/4\pi\varepsilon_0 mc^2$. The quantity $N_e$ can be estimated from the packing density of electrons with a «size» of $r_e$ within a sphere of radius $R_0$ as $N_e \approx (R_0/r_e)^3$. As a result, from expression (1.21), we obtain the estimate $\beta_s^\infty$

$$\beta_s^\infty(R_0) \approx n\frac{\sqrt{2+n^2}}{1+n^2}, \quad n = \frac{R_0}{r_e}, \qquad (1.22)$$

where the quantity $n$ shows how many times the initial radius of the sphere $R_0$ exceeds the classical electron radius $r_e$. Note that $\beta_s^\infty(R_0) < 1$, i.e., $v_s^\infty(R_0) < c$. For the initial distribution (1.14), in both the relativistic (1.17) and classical (1.20) cases, the velocities $v_s^\infty$ and $\mathrm{v}_s^\infty$ are bounded.

In the left panel of Fig. 3, the graphs of the evolution of the ratio $\beta_s(t, R_0) = v_s(t, R_0)/c$ for various spherical layers $R_{0,k}$, $k = 1...8$, found using formula (1.16) with the initial distribution (1.14), are shown. The initial radius satisfy the condition $R_{0,k+1} > R_{0,k}$, $k = 1...7$. The horizontal axis in the left panel of Fig. 3 represents time in relative units, and the vertical axis represents $\beta_s$ is the ratio of the layer's speed to the speed of light. Each curve in the left panel of Fig. 3 has its own asymptote (1.17). Since the initial velocity of all layers is zero, all graphs in the left panel of Fig. 3 start from the origin. The colors of the graphs in the left panel of Fig. 3 and in the left panel of Fig. 1 correspond to the same initial sphere radius $R_{0,k}$. The larger the value of $R_{0,k+1} > R_{0,k}$ (see Fig. 1, left), the higher the limiting value $\beta_s^\infty$ (see the horizontal dashed lines in Fig. 3, left).

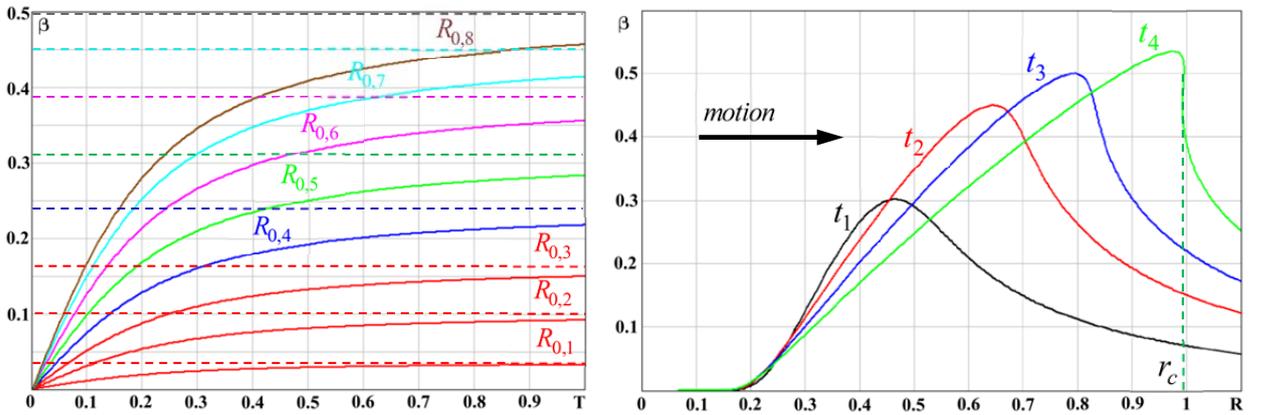

Fig. 3 Graphs of $\beta_s$ for initial distributions (1.14) (left) and (1.15) (right)

The right panel of Fig. 3 shows the coordinate distributions of $\beta_s = v_s/c$ inside the sphere at different times for the log-normal initial distribution (1.15). The distributions $\beta_s$ were obtained using expression (1.16) and the equation of motion (1.10). The horizontal axis in the right panel of Fig. 3 represents the radius values, and the vertical axis represents the values of $\beta_s$. The time moments in the right panel of Fig. 3 satisfy the condition $t_{k+1} > t_k$, $k = 1...3$. The regions with maximum medium velocities gradually shift to larger radius. It can be seen in the right panel of Fig. 3 that the right boundary of the velocity distribution front becomes increasingly steeper over time (compare the black and green graphs). At the moment $t_4$ the velocity distribution (green graph) has a vertical «drop» or gradient catastrophe at the point $R_c$. Consequently, in a small neighborhood of the point $R_c$, there is a larger scatter of velocities. At the point $R_c$, the characteristics intersect (see Fig. 1, right), leading to a shock wave.

Expression (1.10) allows us to construct a formula describing the evolution of the charge density $\rho_s$ up to the time $t < t_c$, when the spherical layers (1.10) do not intersect.

**Theorem 2** *Let the system be at rest initially ( $\dot{R}_0 = 0$ ) with a charge density distribution $\rho_0(r)$, then, for $t < t_c$, the evolution of the charge density $\rho_s[R(t,r),t]$ is described by the expression:*

$$\rho_s[R(t,r),t] = \frac{\rho_0(r)}{\mathcal{P}_s^2(\lambda_s t, \eta_s^2)\left[\mathcal{P}_s(\lambda_s t, \eta_s^2) + r\frac{t\lambda_s' - 2\eta_s \eta_s' \partial_{\eta_s^2}\mathcal{F}_s(\mathcal{P}_s, \eta_s^2)}{\partial_{\mathcal{P}_s}\mathcal{F}_s(\mathcal{P}_s, \eta_s^2)}\right]}, \quad (1.23)$$

*where*

$$2\eta_s(r)\eta_s'(r) = \frac{qr\rho_0(r)}{\varepsilon_0 mc^2} - \frac{\eta_s^2(r)}{r}, \quad (1.24)$$

$$\lambda_s'(r) = \frac{qr^2\rho_0(r) - \varepsilon_0 mc^2 \eta_s^2(r)\{1 + [1+\eta_s^2(r)][2+\eta_s^2(r)]\}}{\varepsilon_0 mc[1+\eta_s^2(r)]^2 r^2 \eta_s(r)\sqrt{2+\eta_s^2(r)}}, \quad (1.25)$$

$$\partial_x \mathcal{F}_s(x,y) = \left[(x-1)\left(x - \frac{y}{y+2}\right)\right]^{-1/2} \frac{1 + (1+y)[x+(x-1)(y+1)]}{(1+y)(2+y)}, \quad (1.26)$$

$$\partial_y \mathcal{F}_s(x,y) = \frac{-1}{(2+y)^2(1+y)}\left\{y\sqrt{\frac{(x-1)(y+2)}{x(y+2)-y}} + \frac{2y+3}{1+y}\operatorname{arcch}[(2+y)x - 1 - y]\right\}. \quad (1.27)$$

The proof of Theorem 2 is given in Appendix A.

**Remark 4** Note that according to the representation (1.10), the equality $\mathcal{P}_s[0, \eta_s^2(R_0)] = 1$, since $R(0, R_0) = R_0$. Consequently, $\mathcal{F}_s[1, \eta_s^2(r)] = 0$, which agrees with expression (1.9). Thus, at the initial moment, expression (1.23) reduces to the distribution $\rho_0(r)$.

Note that expression (1.16) can be rewritten using the derivative (1.26). Indeed, from the definition of the inverse function $\mathcal{P}_s(\mathcal{F}_s(x,y), y) = x$ it follows that $\partial_{\mathcal{F}_s}\mathcal{P}(\mathcal{F}_s, y)_s \partial_x \mathcal{F}_s(x,y) = 1$. As a result, the time derivative $v_s = \dot{R}(t, R_0)$ of the equation of motion (1.10) takes the form:

$$v_s(t, R_0) = \frac{R_0 \lambda_s(R_0)}{\partial_{\mathcal{P}_s} \mathcal{F}_s \{\mathcal{P}_s[\lambda_s(R_0)t, \eta_s^2(R_0)], \eta_s^2(R_0)\}}. \tag{1.28}$$

In the limiting non-relativistic case $c \to +\infty$, expression (1.28) transforms into (1.19), since $\partial_{\mathcal{P}_s}\mathcal{F}_s(\mathcal{P}_s, \eta_s^2) \to \sqrt{\mathcal{P}_s/(\mathcal{P}_s - 1)}$.

**Corollary 4** *In the non-relativistic limit $c \to +\infty$, the expression (1.23) for the density $\rho_s$ takes the form*

$$\lim_{c \to +\infty} \rho_s[R(t,r), t] = \rho_s[\mathrm{R}(t,r), t] = \tag{1.29}$$

$$= \frac{\rho_0(r)}{P_s^2[\lambda_s(r)t]\left\{P_s[\lambda_s(r)t] + t\sqrt{\frac{P_s[\lambda_s(r)t]-1}{P_s[\lambda_s(r)t]}}\left[\frac{q\rho_0(r)}{m\varepsilon_0\lambda_s(r)} - \frac{3}{2}\lambda_s(r)\right]\right\}},$$

*where, according to (1.13), the condition $P_s(0) = 1$ is satisfied.*

The proof of Corollary 4 is given in Appendix A.

Figure 4 shows graphs of the charge density distributions (1.23) at different times for the two initial distributions (1.14) on the left and (1.15) on the right. The horizontal axis represents the radius in relative units, and the vertical axis represents the density $\rho_s$. Let us first consider the graphs in the left panel of Fig. 4. At the initial time $t = 0$, the charge density is constant inside the sphere (1.14) (black graph).

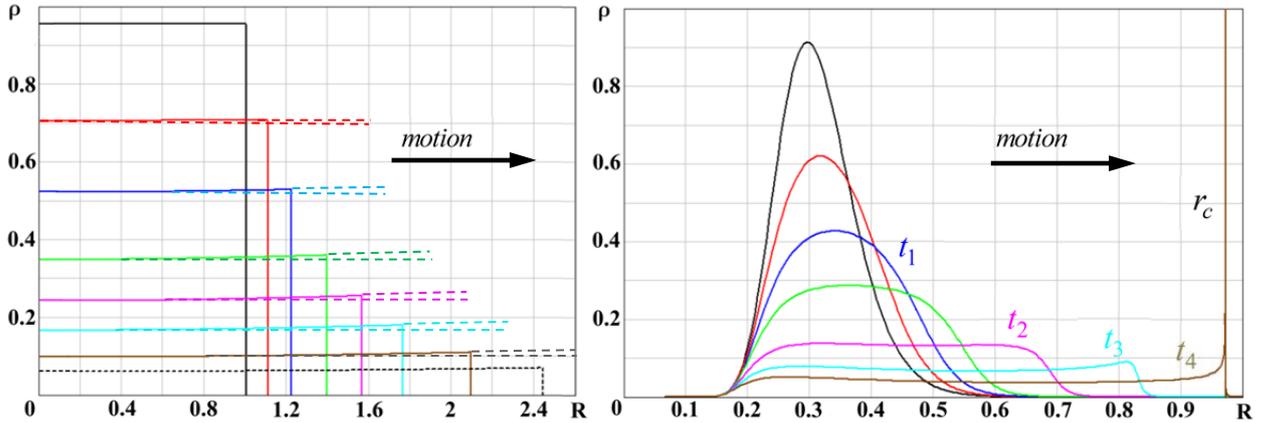

Fig. 4 Evolution for the initial distributions (1.14) on the left and (1.15) on the right

At subsequent times, according to the characteristic equations (1.10) (see Fig. 1, left), the sphere expands, and as a consequence, the charge density is distributed almost uniformly throughout its volume (see Fig. 4, left). In the non-relativistic case, according to (1.29), the density would be homogeneous at each moment in time:

$$\rho_s[\mathrm{R}(t,r), t] = \frac{\rho_0(r)}{P_s^3[\lambda_s(r)t]}, \tag{1.30}$$

where it is taken into account that $\lambda'_s = 0$, i.e., $\lambda_s = \sqrt{2q\rho_0/3\varepsilon_0 m} = const$. In the left panel of Fig. 4, the dashed lines show the slope that arises in the density distribution during the expansion of the sphere.

Thus, in the relativistic case, a redistribution of the charge density occurs, characterized by an accumulation of charge in the outer layers of the sphere. This effect has an intuitive relativistic interpretation. The point is that the maximum velocity is reached in the outer layers of the sphere (see Fig. 3, left, and Fig. 1, left). Consequently, the effect of the increase in relativistic mass $m_r = m\gamma$ is most significant in the outer layers, which leads to a deceleration of their motion. This deceleration creates a charge accumulation effect, which is observed in Fig. 4, left.

Let us consider the evolution of the charge density (see Fig. 4, right) for the initial distribution (1.15). The black curve in Fig. 4, right, shows the radial charge density distribution at the initial moment. Essentially, the density distribution resembles a charged spherical shell. During the system's expansion, the width of the spherical shell gradually increases, and its density decreases (see Fig. 4, right). At the same time, according to the velocity graphs in Fig. 3, right, the maximum charge flow velocity is located inside the shell. The time moments $t_k, k = 1...4$ in Fig. 4, right, correspond to the time moments in Fig. 3, right. Thus, unlike the previous case, the maximum effect on the relativistic mass occurs not at the edge of the distribution but at its «center». However, this situation is only present at the beginning of the expansion process. It can be seen in Fig. 3, right, that the maximum charge flow velocity gradually shifts towards the edge of the distribution. Consequently, the effect on the relativistic mass $m_r$ shifts as well. As a result, a high density arises at the edge of the distribution (see Fig. 4, right, brown graph). At the moment $t_4 \approx t_c$ in Fig. 3, right, a gradient catastrophe occurs, and in Fig. 4, right, the charge density $\rho_s$ begins to grow unboundedly due to the intersection of characteristics in Fig. 1, right. Note that in the absence of relativism, expression (1.29) with the initial distribution (1.15) would also lead to a shock wave. Recall that the relativistic effect, according to Fig. 2, causes a delay in the shock wave effect.

Knowing the charge density distributions at each moment in time $\rho_s(r,t)$, one can construct the evolution of the charge distribution itself (1.1).

**Lemma 1** *Let $\rho_s(r,t)$ be the solution to the space charge problem from Theorem 2. Then the charge evolution function (1.1) $q\,\mathrm{N}_s(r,t) = 4\pi \int_0^r \rho_s(x,t) x^2 dx$ satisfies the equation:*

$$\frac{d}{dt}\mathrm{N}_s(r,t) = \left[\frac{\partial}{\partial t} + \langle v \rangle(r,t) \frac{\partial}{\partial r}\right] \mathrm{N}_s(r,t) = 0, \quad (1.31)$$

$$\langle v \rangle \bigl[R(t,R_0),t\bigr] \stackrel{\text{def}}{=} c\beta_s(t,R_0), \quad (1.32)$$

*which has a solution in the form of the characteristic (1.10) $\mathrm{N}_s\bigl[R(t,R_0),t\bigr] = const$.*

The proof of Lemma 1 is given in Appendix A.

Figure 5 shows the distributions of $\mathrm{N}_s(r,t)$ for case (1.14) on the left and case (1.15) on the right. The horizontal axes represent time $t$ and radius $r$, and the vertical axis represents the solution $\mathrm{N}_s(r,t)$ itself. The red lines in Fig. 5 show the graphs of the characteristics (1.10) (compare with Fig. 1). According to Lemma 1, the solution to equation (1.31) remains constant along these characteristics, i.e., $N_s = const$. Thus, the initial charge (1.1) is conserved within each spherical layer. Essentially, the characteristics (1.10) are the level lines for the function $N_s(r,t)$.

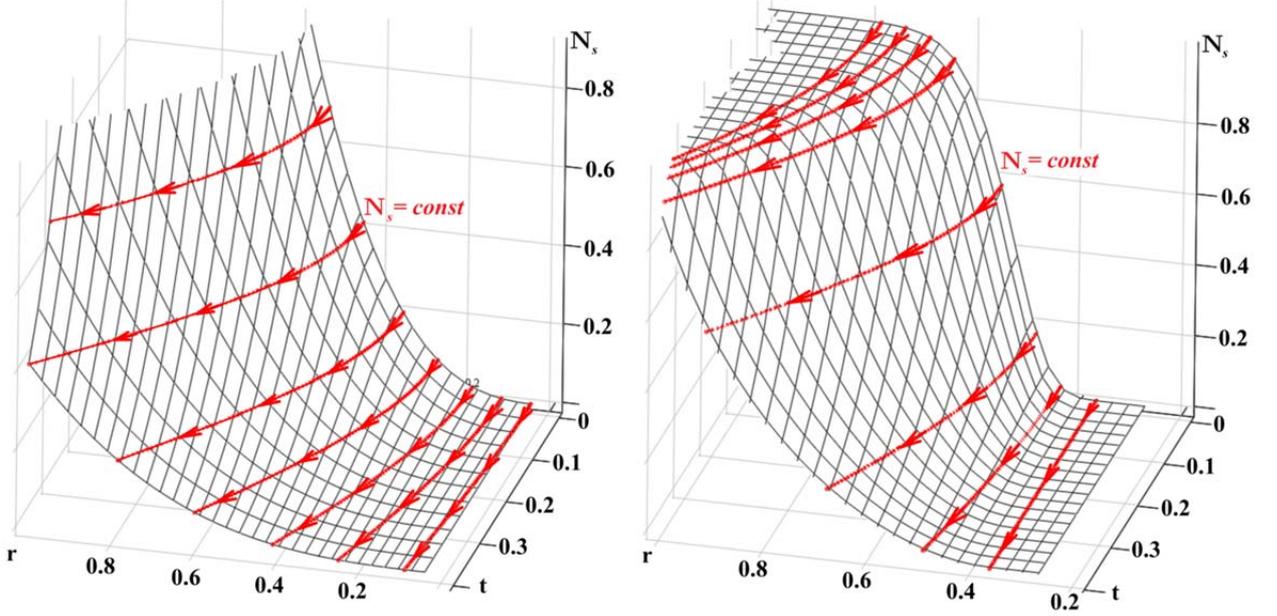

Fig. 5 The evolution of $N_s$ for initial distributions (1.14) on the left and (1.15) on the right

## 1.2 Cylindrically symmetric charge density distribution

Let us consider the formulation of the space charge problem with cylindrical symmetry. It is assumed that the charge density is homogeneous in the longitudinal direction, and there is azimuthal angular symmetry in the circular cross-section of the cylinder. Inhomogeneity in the density distribution is only possible in the radial direction. In the longitudinal (axial) direction, the cylindrical region has an infinite length. Due to this symmetry, instead of the volumetric charge density $\rho$, we will consider the charge density distribution in the circular cross-section of the cylinder $\rho_c = \rho \ell$, where $\ell$ corresponds to the height of an elementary cylindrical layer. By analogy with the spherically symmetric distribution (1.1)-(1.3), the Cauchy problem takes the form

$$\frac{d}{dt}\frac{\dot{R}}{\sqrt{1-\dot{R}^2/c^2}} = \frac{\vartheta_c^2}{R}, \quad R(0) = R_0, \quad \dot{R}(0) = 0, \quad \vartheta_c^2 = \frac{q^2 N_c(R_0)}{2\pi\varepsilon_0 m \ell}, \qquad (1.33)$$

where

$$qN_c(R_0) = 2\pi \int_0^{R_0} \rho_c(x,0) x dx = 2\pi \int_0^{R(t,R_0)} \rho_c(x,t) x dx = q\, N_c[R(t,R_0),t].$$

**Theorem 3** *The Cauchy problem (1.33) has a solution $R(t, R_0)$ defined parametrically*

$$\mathcal{F}_c\left[\frac{R(t,R_0)}{R_0}, \bar{\beta}_c^2(R_0)\right] = \lambda_c(R_0)t, \qquad (1.34)$$

$$\mathcal{F}_c(x,y) \stackrel{def}{=} \sqrt{2} \int_0^{\sqrt{\ln x}} \frac{1+yz^2}{\sqrt{2+yz^2}} e^{z^2} dz, \quad \lambda_c(r) \stackrel{def}{=} \frac{\vartheta_c(r)}{r\sqrt{2}}, \qquad (1.35)$$

where $\bar{\beta}_c(R_0) = \vartheta_c(R_0)/c$. In the classical limit $c \to +\infty$, equation (1.34) takes the form:

$$F_c\left[\frac{R(t,R_0)}{R_0}\right] = \lambda_c(R_0)t, \qquad F_c(x) \stackrel{\text{def}}{=} \lim_{c \to +\infty} \mathcal{F}_c(x,y) = \int_0^{\sqrt{\ln x}} e^{z^2} dz, \qquad (1.36)$$

where $P_c(F_c) = x$ and $R(t,R_0) = R_0 P_c[\lambda_c(R_0)t]$.

The proof of Theorem 3 is given in Appendix A.

**Theorem 4** *Let the evolution of charged cylindrical layers be described by equation (1.34), and at the initial time, it has a charge density distribution $\rho_0(r)$, then the charge density function $\rho_c[R(t,r),t]$ for $t < t_c$ is given by:*

$$\rho_c[R(t,r),t] = \frac{\rho_0(r)}{\mathcal{P}_c(\lambda_c t, \bar{\beta}_c^2)\left[\mathcal{P}_c(\lambda_c t, \bar{\beta}_c^2) + r\dfrac{t\lambda_c' - 2\bar{\beta}_c\bar{\beta}_c' \partial_{\bar{\beta}_c^2}\mathcal{F}_c(\mathcal{P}_c, \bar{\beta}_c^2)}{\partial_{\mathcal{P}_c}\mathcal{F}_c(\mathcal{P}_c, \bar{\beta}_c^2)}\right]}, \qquad (1.37)$$

where

$$2\bar{\beta}_c(r)\bar{\beta}_c'(r) = \frac{qr\rho_0(r)}{\varepsilon_0 \ell mc^2}, \qquad \lambda_c'(r) = \frac{\lambda_c(r)}{r}\left[\frac{\rho_0(r)\pi r^2}{qN_c(r)} - 1\right], \qquad (1.38)$$

$$\partial_x \mathcal{F}_c(x,y) = \frac{1 + y\ln x}{\sqrt{2\ln x(2 + y\ln x)}}, \qquad \partial_y \mathcal{F}_c(x,y) = \frac{1}{\sqrt{2}}\int_0^{\sqrt{\ln x}} \frac{3 + yz^2}{(2 + yz^2)^{3/2}} z^2 e^{z^2} dz, \qquad (1.40)$$

and the function $\mathcal{P}_c$ is the inverse of the function $\mathcal{F}_c(x,y)$ with respect to the variable $x$ ($x \geq 1$), i.e., $R(t,R_0) = R_0 \mathcal{P}_c[\lambda_c(R_0)t, \bar{\beta}_c^2(R_0)]$.

The proof of Theorem 4 is given in Appendix A.

**Corollary 5** *It follows from Theorem 4 that the expression for the charge flow velocity ($\beta = v/c$) is:*

$$\beta_c(t,R_0) = \bar{\beta}_c(R_0)\frac{\sqrt{\kappa_c(t,R_0)[2 + \bar{\beta}_c^2(R_0)\kappa_c(t,R_0)]}}{1 + \bar{\beta}_c^2(R_0)\kappa_c(t,R_0)}, \qquad (1.41)$$

$$\kappa_c(t,R_0) \stackrel{\text{def}}{=} \ln \mathcal{P}_c[\lambda_c(R_0)t, \bar{\beta}_c^2(R_0)],$$

with the asymptotics:

$$\beta_c^\infty(R_0) = \lim_{t \to +\infty} \beta_c(t,R_0) = 1. \qquad (1.42)$$

In the non-relativistic limit $c \to +\infty$, the expressions for the charge density (1.37) and the velocity (1.41) transform into the following, respectively:

$$\rho_c\left[R(t,r),t\right] \stackrel{def}{=} \lim_{c\to+\infty} \rho_c\left[R(t,r),t\right] = \qquad (1.43)$$

$$= \frac{\rho_0(r)}{P_c\left[\lambda_c(r)t\right]\left\{P_c\left[\lambda_c(r)t\right]+2t\sqrt{\ln P_c\left[\lambda_c(r)t\right]}\left[\frac{q\rho_0(r)}{4\varepsilon_0 m\ell\lambda_c(r)}-\lambda_c(r)\right]\right\}},$$

$$v_c(t,R_0) \stackrel{def}{=} \lim_{c\to+\infty} c\beta_c(t,R_0) = 2R_0\lambda_c(R_0)\sqrt{\ln P_c\left[\lambda_c(R_0)t\right]}, \quad v_c^\infty(R_0) \stackrel{def}{=} \lim_{t\to+\infty} v_c(t,R_0) = +\infty. \ (1.44)$$

The proof of Corollary 5 is given in Appendix A.

Let us demonstrate the behavior of the velocities of cylindrical layers (1.41) and (1.44) at large times $t$. It seems natural to consider a homogeneous initial charge density distribution of type (1.14), as in this case the characteristics (1.34) do not intersect. Figure 6 shows the graphs of the evolution of the functions $\beta_c(t,R_0)$ (1.41) for the relativistic case (solid lines) and the graphs of $v_c(t,R_0)/c$ for the classical case (dashed lines). Each graph corresponds to a cylindrical layer with its own initial radius $R_{0,j}$, satisfying the condition $R_{0,j+1} > R_{0,j}$. The $R_{0,j}$ values are the same for the classical and relativistic cases, so the corresponding graphs are denoted by the same color in Fig. 6. In Fig. 6, two distinct groups of curves (classical and relativistic cases) are clearly visible, separated by the dashed horizontal relativistic asymptote $\beta_c^\infty(R_0) = 1$ (1.42).

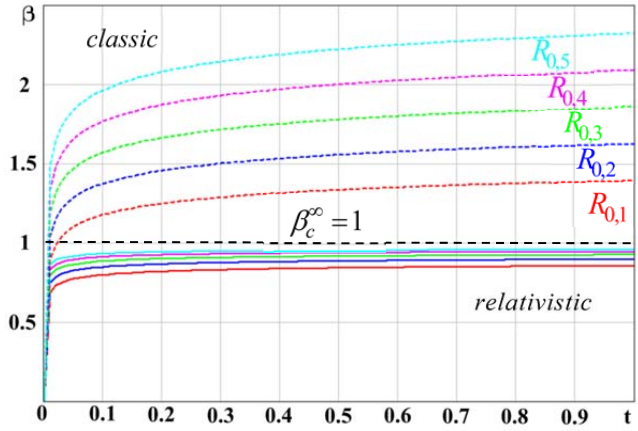

Fig. 6 The evolution of function $\beta_c$ and $v_c/c$

It can be seen in Fig. 6 that in the relativistic case, the functions $\beta_c(t,R_{0,k})$ with $k=1...5$ according to (1.42) tend to the same value $\beta_c^\infty(R_0) = 1$. During the evolution, the velocity of the outer layer (i.e., $R_{0,5}$) exceeds the velocity of the inner layers ($R_{0,4}, R_{0,3}, R_{0,2}, R_{0,1}$). Nevertheless, over time, the difference between the velocities of the cylindrical layers decreases, and their values asymptotically approach the speed of light. In the classical case (dashed lines in Fig. 6), the velocities of the cylindrical layers according to (1.44) grow unboundedly. The difference between the velocities of adjacent layers increases monotonically.

Note that when considering spherical layers (1.16)-(1.17), each layer in both the relativistic and classical cases had its own limiting velocity value (1.17) and (1.20) (see Fig. 3, left). In a system with cylindrical symmetry, finite velocities exist only in the relativistic case (1.42).

The evolution of the charge density $\rho_c$ or $\rho_c$ of cylindrical layers exhibits behavior similar to that of spherical layers, both in the relativistic (1.37) and classical (1.43) cases. For an inhomogeneous initial log-normal distribution of type (1.15), a shock wave effect is observed, similar to the effect depicted in Fig. 4, right. As in the spherically symmetric case, accounting for relativism leads to a delay in the shock wave effect (see Fig. 2) for the characteristics (1.34). The formulation of Lemma 1 can be generalized to the function $N_c$ (1.33), resulting in equation (1.31) $dN_c/dt = 0$ for it. In this case, the equations of motion (1.34) represent the characteristics along which $N_c = const$.

## §2 Gravitational case

The consideration of gravitational problems requires the involvement of General Relativity (GR), which accounts for the curvature of the space-time metric $g_{\nu\mu}$ under the influence of masses distributed in space, while the equations of motion are essentially the geodesic equations. Accounting for the self-consistency condition significantly complicates the solution of such a problem within the framework of GR.

This section considers a simplified model of «quasi-relativism», which uses the Newtonian formula for the force of attraction, while the momentum in the equation of motion is relativistic. The Newtonian approximation is acceptable for orders of magnitude where $Gm/c^2 \ll 1$.

### 2.1 Spherically symmetric mass density distribution

By analogy with the electromagnetic case from §1 et us consider a spherically symmetric system with gravitational interaction. We take the following as the equation of motion for a spherical layer (see (1.2)):

$$\frac{d}{dt}\frac{\dot{R}}{\sqrt{1-\dot{R}^2/c^2}} = -\frac{\vartheta_s^2}{R^2}, \quad \vartheta_s^2(R_0) \overset{\text{def}}{=} \frac{1}{2} r_g c^2 N_s(R_0), \quad r_g = \frac{2Gm}{c^2}, \qquad (2.1)$$

$$mN_s(R_0) = 4\pi \int_0^{R_0} \rho_s(x,0) x^2 dx = 4\pi \int_0^{R(t,R_0)} \rho_s(x,t) x^2 dx \overset{\text{def}}{=} m\,\mathrm{N}_s\left[R(t,R_0),t\right], \qquad (2.2)$$

where $\rho_s(r,t) = mf_1(r,t)$; $G$ is the gravitational constant, and the function $mN_s(R_0) = M_s(R_0)$ specifies the mass contained within a sphere of radius $R_0$. The quantity defines the gravitational radius (Schwarzschild radius). Compare expressions (2.1) and (1.2) for $\vartheta_s^2$. The minus sign on the right-hand side of equation (2.1) indicates the process of compression under the action of the gravitational force of attraction.

**Theorem 5** *The Cauchy problem (1.2) for equation (2.1) has a solution $R(t,R_0)$*

$$\mathcal{F}_s\left[\frac{R(t,R_0)}{R_0}, \eta_s^2(R_0)\right] = \lambda_s(R_0)t, \qquad (2.3)$$

*where*

$$\mathcal{F}_s(x,y) \overset{\text{def}}{=} \sqrt{(1-x)\left(x+\frac{y}{2-y}\right)} + \frac{\arccos\left[(2-y)x+y-1\right]}{(1-y)(2-y)}, \qquad (2.4)$$

$$\lambda_s(R_0) \overset{\text{def}}{=} \frac{c\eta_s(R_0)\sqrt{2-\eta_s^2(R_0)}}{R_0\left[1-\eta_s^2(R_0)\right]}. \qquad (2.5)$$

*In the classical limit, the characteristics (2.3) take the form*

$$\lim_{c\to+\infty} \mathcal{F}_s(x,\eta_s^2) = \sqrt{x(1-x)} + \arccos\sqrt{x} = F_s(x), \quad \lim_{c\to+\infty} \lambda_s(r) = \frac{\vartheta_s(r)\sqrt{2}}{r^{3/2}} = \lambda_s(r), \qquad (2.6)$$

*where the representations (1.10) and (1.13) are valid.*

The proof of Theorem 5 is given in Appendix B.

**Remark 5**

Note that the equations of motion (1.2) and (2.1) differ only in the signs of their right-hand sides. On the one hand, such a «minor» difference leads to a certain similarity in the structure of their solutions. Indeed, compare expressions (1.7)-(1.9) and expressions (2.3)-(2.5). On the other hand, the solutions differ substantially, as they correspond to diametrically opposite physical processes: expansion (electromagnetic interaction) and compression (gravitational interaction).

In the left and right panels of Fig. 7, the graphs of the characteristics (2.3) are shown for the two initial mass density distributions of the form (1.14) and (1.15), respectively. The force of gravitational attraction leads to the compression of the spherical layers, as shown by the arrows pointing towards the center in Fig. 7 (compare with Fig. 1). In the left panel of Fig. 7, the red solid lines show the characteristics accounting for relativism (2.3), and the blue dashed lines show the characteristics for the classical limit (2.6). It can be seen that in the classical limit (blue dashed lines), the characteristics intersect at the center of the sphere, while accounting for relativism (red solid line) leads to the disappearance of the «shock wave effect». In the case of the initial log-normal distribution (1.15), the intersection of characteristics at a certain inner radius $R_c$ is observed in both the classical (2.6) and relativistic (2.4) cases (see Fig. 7, right). Accounting for relativism in this case leads to a delay of the «shock wave», meaning the intersection of characteristics occurs at a smaller radius $R_c$. A similar «slowing down» effect was observed in Fig. 2 for the electric field.

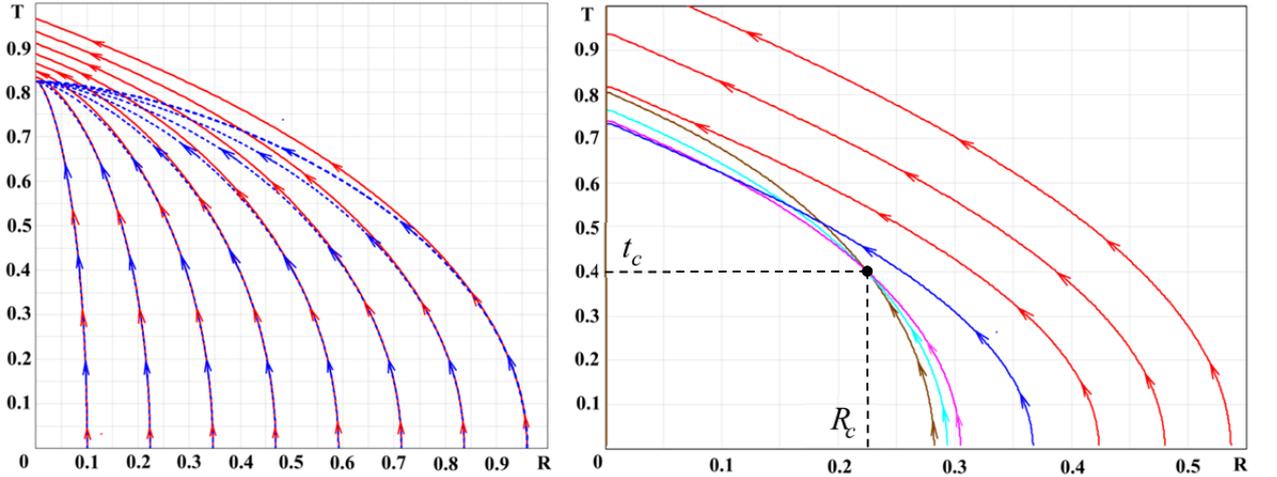

Fig. 7 Graphs of characteristics (2.3) for distributions (1.14)-(1.15)

**Theorem 6** *The evolution of the mass density function $\rho_s$ for the characteristics (2.7) for $t < t_c$ is described by expression (1.23), in which*

$$2\eta_s(r)\eta_s'(r) = \frac{4\pi G r \rho_0(r)}{c^2} - \frac{\eta_s^2(r)}{r}, \qquad (2.7)$$

$$\lambda_s'(r) = \frac{4\pi G r^2 \rho_0(r) - c^2 \eta_s^2(r)\{1 + [1 - \eta_s^2(r)][2 - \eta_s^2(r)]\}}{c[1 - \eta_s^2(r)]^2 r^2 \eta_s(r)\sqrt{2 - \eta_s^2(r)}}, \qquad (2.8)$$

$$\partial_x \mathcal{F}_s(x,y) = -\left[(1-x)\left(x + \frac{y}{2-y}\right)\right]^{-1/2} \frac{1 + (1-y)[x - (1-x)(1-y)]}{(1-y)(2-y)}, \qquad (2.9)$$

$$\partial_y \mathcal{F}_s(x,y) = \frac{1}{(2-y)^2(1-y)} \left\{ -y\sqrt{\frac{(1-x)(2-y)}{x(2-y)+y}} + \frac{3-2y}{1-y} \arccos\left[(2-y)x-1+y\right] \right\}, \qquad (2.10)$$

*where the quasi-relativism condition $y<1$ is assumed to be satisfied. In the classical limit $c \to +\infty$, the next expression is valid:*

$$\lim_{c \to +\infty} \rho_s\left[R(t,r),t\right] = \rho_s\left[\mathrm{R}(t,r),t\right] = \qquad (2.11)$$

$$= \frac{\rho_0(r)}{P_s^2\left[\lambda_s(r)t\right]\left\{P_s\left[\lambda_s(r)t\right] - t\sqrt{\frac{1-P_s\left[\lambda_s(r)t\right]}{P_s\left[\lambda_s(r)t\right]}}\left[\frac{4\pi G \rho_0(r)}{\lambda_s(r)} - \frac{3}{2}\lambda_s(r)\right]\right\}}.$$

The proof of Theorem 6 is given in Appendix B.

Figures 8 and 9 show the evolution of the mass density $\rho_s$ for the initial distribution (1.14) and the log-normal distribution (1.15), respectively. The black arrows in Figs. 8-9 indicate the direction of mass motion (towards the center of the sphere). The left panel of Fig. 8 shows the distributions $\rho_s$ for the initial time moments. The graph numbers correspond to the time moments $t_n$, where $\tau$ is a certain time step.

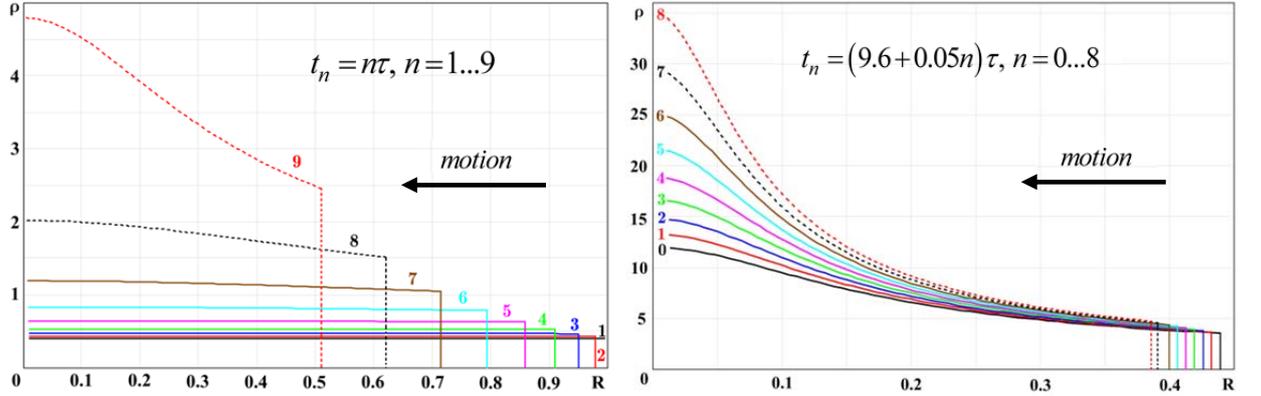

Fig. 8 Evolution of mass density with initial distribution (1.14)

Note that in the non-relativistic case (2.11) with (1.14) for $\rho_s$ expression (1.30) is valid. According to (1.30), at each moment in time, the mass density $\rho_s$ is constant with respect to the radius. It can be seen in the left panel of Fig. 8 that at the initial times, the behavior of $\rho_s$ is close in character to (1.30). Indeed, at $t_1,...,t_6$, the velocities of the spherical layers are small, so the relativistic effect is negligible. As a result, the density distributions at $t_1,...,t_6$ in the left panel of Fig. 8 appear homogeneous inside the sphere. As the gravitational compression of the sphere proceeds, the velocities of the layers increase, and the relativistic effect becomes more substantial. It can be seen in the left panel of Fig. 8 that at time moments $t_7,...,t_9$ a strong slope appears in the distribution of $\rho_s$. A similar picture was observed earlier in the problem with electric interaction in Fig. 4, left. The right panel of Fig. 8 shows the further evolution of the density $\rho_s$, which is distinguished by a strongly nonlinear character. The graphs in the right panel of Fig. 8 are plotted over significantly smaller time intervals than the analogous ones in the left panel of Fig. 8.

Note that the density evolutions $\rho_s$ in Fig. 8 correspond to the characteristics in the left panel of Fig. 7. At the initial times $t_1,..,t_6$ the behavior is close to the non-relativistic case for both the characteristics in Fig. 7, left, and the density $\rho_s$ in Fig. 8. Indeed, the blue and red characteristics in Fig. 7, left, almost coincide. The difference arises at later times (divergence of the blue and red lines in Fig. 7, left).

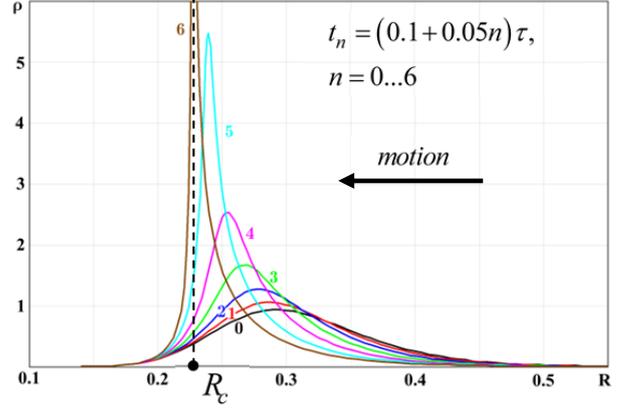

Fig. 9 Evolution of mass density with initial distribution (1.15)

The evolution of the density $\rho_s$ in Fig. 9 has a significant difference from that in Fig. 8. The inhomogeneous initial distribution $\rho_0$ (1.15) according to Fig. 7, right, leads to the intersection of characteristics at an inner radius $R_c$. From a mathematical point of view, the density $\rho_s$ grows infinitely at the point $R_c$ (see Fig. 9). As a result, a spherical shell of radius $R_c$ with a large mass density is effectively formed.

**2.2 Cylindrically symmetric mass density distribution**
In the case of gravitational interaction, the formulation (1.33) takes the form:

$$\frac{d}{dt}\frac{\dot{R}}{\sqrt{1-\dot{R}^2/c^2}} = -\frac{\vartheta_c^2}{R}, \quad R(0)=R_0, \; \dot{R}(0)=0, \qquad \vartheta_c^2 = \frac{2G}{\ell}M_c(R_0), \qquad (2.12)$$

where

$$M_c(R_0) = mN_c(R_0) = 2\pi \int_0^{R_0} \rho_c(x,0) x dx = 2\pi \int_0^{R(t,R_0)} \rho_c(x,t) x dx = m\,\mathrm{N}_c[R(t,R_0),t].$$

**Theorem 7.** *The Cauchy problem (2.12) has a solution $R(t,R_0)$ in the form of characteristics (1.34), for which:*

$$\mathcal{F}_c(x,y) \stackrel{def}{=} \sqrt{2} \int_0^{\sqrt{\ln(1/x)}} \frac{1+yz^2}{\sqrt{2+yz^2}} e^{-z^2} dz, \qquad (2.13)$$

$$F_c(x) \stackrel{def}{=} \lim_{c \to +\infty} \mathcal{F}_c(x,y) = \frac{\sqrt{\pi}}{2}\mathrm{erf}\left(\sqrt{\ln\frac{1}{x}}\right), \qquad (2.14)$$

*where the function* $\mathrm{erf}$ *is the error function and $x \le 1$. The density function $\rho_c$ has the form (1.37) with derivatives:*

$$\bar{\beta}_c(r)\bar{\beta}'_c(r) = \frac{2\pi G}{\ell c^2} r\rho_0(r), \quad \lambda'_c(r) = \frac{1}{r^2}\sqrt{\frac{G}{\ell}M_c(r)}\left[\frac{\pi r^2 \rho_0(r)}{M_c(r)} - 1\right], \qquad (2.15)$$

$$\partial_x \mathcal{F}_c(x,y) = \frac{y\ln x - 1}{\sqrt{2\ln x}(y\ln x - 2)}, \quad \partial_y \mathcal{F}_c(x,y) = \frac{1}{\sqrt{2}} \int_0^{\sqrt{\ln(1/x)}} \frac{(3+yz^2)z^2}{(2+yz^2)^{3/2}} e^{-z^2} dz. \quad (2.16)$$

*In the non-relativistic limit $c \to +\infty$, the expression for the mass density $\rho_c$ becomes:*

$$\rho_c[R(t,r),t] \stackrel{def}{=} \lim_{c \to +\infty} \rho_c[R(t,r),t] = \qquad (2.17)$$

$$= \frac{\rho_0(r)}{P_c[\lambda_c(r)t]\left\{P_c[\lambda_c(r)t] - 2t\sqrt{\ln\frac{1}{P_c[\lambda_c(r)t]}}\left[\frac{\pi G \rho_0(r)}{\ell \lambda_c(r)} - \lambda_c(r)\right]\right\}}.$$

The proof of Theorem 7 is given in Appendix B.

Let us compare the gravitational collapse times for systems with spherical and cylindrical symmetry. If the initial mass density distribution $\rho_0(r)$ is inhomogeneous, then an intersection of characteristics can occur at an inner radius $R_c \neq 0$ (see Fig. 9). Therefore, it is necessary to take a homogeneous initial distribution (1.14). In the relativistic case, each layer will reach the origin at its own time (see Fig. 8, right, and Fig. 7, left), so let us consider the non-relativistic case. The collapse time for the spherical $T_s$ and cylindrical $T_c$ cases is obtained from the characteristic equations (2.3) and (2.24), respectively:

$$T_s = \frac{F_s(0)}{\lambda_s(R_0)} = \frac{1}{4}\sqrt{\frac{3\pi}{2G\rho_0}}, \quad T_c = \frac{F_c(0)}{\lambda_c(R_0)} = \frac{1}{2\sqrt{G\rho_0}}, \quad \frac{T_s}{T_c} = \sqrt{\frac{3}{8}\pi}, \quad (2.18)$$

where it is taken into account that the density $\rho_0/\ell$ in the cylindrical case corresponds to the density in the spherically symmetric case. As can be seen from expressions (2.18), the times $T_s$ and $T_c$ do not depend on the initial radius of the layer $R_0$ (see Fig. 7, left, blue curves). In this sense, gravitational collapse is similar to the shock wave effect. From relation (2.8), it follows that the radial collapse times satisfy the condition $T_s > T_c$.

### §3 Discussion

The Wigner-Vlasov formalism relies on the Vlasov chain of equations for the distribution functions. This work considers only the first two equations (i.1)-(i.2). As noted in the Introduction, the physical nature of the functions $f_n$ can be different: probability density, charge or mass density, particle distribution function.

From a mathematical standpoint, relations (i.1)-(i.23) allow one to determine the main physical characteristics from the distribution function. For example, suppose the function $f_2$, satisfying equation (i.2) is known. Then, using expressions (i.3), one finds $f_1$ and $\langle \vec{v} \rangle$, satisfying the first Vlasov equation (i.1). The average velocity field $\langle \vec{v} \rangle$, up to a gauge (i.12), is decomposed into potential $-2\alpha_1 \nabla_r \varphi$ and vortex $\alpha_2 \vec{A}_\Psi$ components (i.8) by Helmholtz's theorem. As a result, one can construct the wave function $\Psi = \pm\sqrt{f_1}\exp(i\varphi)$, satisfying the Schrödinger-type equation (i.9). Since $\Psi$ and $\vec{A}_\Psi$ are known, the potential $U$ is found from equation (i.9), and the quantum potential Q is obtained directly from $|\Psi|$ via (i.10). Consequently, the Hamilton-Jacobi-type equation (i.10) is also satisfied. The fields $\vec{E}_\Psi, \vec{B}_\Psi$ are determined by formulas (i.11), and the equation of motion is satisfied.

Note that, from a mathematical point of view, the described sequence of actions does not depend on the type of interaction (electromagnetic or gravitational). Within the framework of «quasi-relativism», Newtonian gravity is described by potential theory. Consequently, the law of mass conservation (continuity equation (i.1)) and Helmholtz's theorem (i.7) are valid. Therefore, for the solutions from §2 analogs of the Schrödinger-type equation (i.9), the Hamilton-Jacobi-type equation (i.10), and the equation of motion (i.11) will be valid, where the constant $\alpha_2$ can be taken as $\alpha_2 = -\sqrt{4\pi\varepsilon_0 G}$.

Another question is the fulfillment of the self-consistency condition (i.17), which leads to the construction of the field equations (i.19). On the one hand, the solutions obtained in §1-2, are self-consistent. Indeed, the electric and gravitational fields, under the influence of which the charge and mass density distributions evolve, are the fields of the system itself. There are no other external fields. Therefore, condition (i.17) must be satisfied. On the other hand, relativistic equations of motion (1.2), (1.33) and (2.1), (2.12), which differ from equation (i.11) with the classical momentum $\langle\vec{p}\rangle = m\langle\vec{v}\rangle$, were used to obtain the solutions.

To answer the question of self-consistency, let us consider the relativistic and non-relativistic solutions separately. In the non-relativistic limits $c \to \infty$, the solutions (1.19), (1.29) or (1.43), (1.44) will satisfy the equation of motion (i.11) with the corresponding self-consistent fields. The relativistic solutions (1.16), (1.23) and (1.37), (1.41) also satisfy equation (i.11), but with different effective fields $\vec{E}''_\Psi, \vec{B}''_\Psi$

$$\frac{d}{dt}\langle\vec{v}\rangle = -\alpha_2\left(\vec{E}''_\Psi + \langle\vec{v}\rangle \times \vec{B}''_\Psi\right), \quad \gamma\vec{E}''_\Psi = \vec{E}' + \frac{\dot\gamma}{\alpha_2}\langle\vec{v}\rangle, \quad \gamma\vec{B}''_\Psi = \vec{B}', \tag{3.1}$$

$$\text{div}_r \vec{E}' = \rho/\varepsilon_0 \neq \text{div}_r \vec{E}''_\Psi, \tag{3.2}$$

where, equation (3.1) is obtained from the formal (see Introduction) relativistic equation

$$\frac{d}{dt}\left[\gamma\langle\vec{v}\rangle\right] = -\alpha_2\left(\vec{E}' + \langle\vec{v}\rangle \times \vec{B}'\right). \tag{3.3}$$

Thus, the fields $\vec{E}', \vec{B}'$ satisfy all four Maxwell's equations with the 4-current $J^\nu = \rho(c, \langle\vec{v}\rangle)$, while the fields $\vec{E}''_\Psi, \vec{B}''_\Psi$ do not. In the relativistic case, the gauge (i.12) imposes a condition on the phase $\varphi$ of the wave function (see Appendix B):

$$\Box S = -K, \quad S = \hbar\varphi, \quad K \stackrel{\text{def}}{=} \text{div}_r\langle\vec{p}\rangle + \frac{1}{c^2}\frac{\partial}{\partial t}\frac{\langle p\rangle^2}{2m}, \tag{3.4}$$

where S is the action. The right-hand side of (3.4) is similar to a gauge condition on the kinetic part of the system. The solution to equation (3.4) can be found via the Liénard-Wiechert potential. Knowing the phase $\varphi$ from the decomposition (i.7), the field $\vec{A}_\Psi$ is determined. Despite the fact that when solving relativistic problems in §1-2 $\text{curl}_r \vec{A}_\Psi = \vec{0}$, the field itself $\vec{A}_\Psi \neq \vec{0}$ due to gauge (i.12) with time dependent potential V. It follows that the field $\vec{A}_\Psi$ is vector potential with a single radial component. In the non-relativistic case ($c \to \infty$), the gauge (i.12) transforms into the Coulomb gauge $\text{div}_r \vec{A}_\Psi = \vec{0}$. As a result, in the decomposition (i.7), one can use the representation $\langle \vec{v} \rangle = -\alpha_1 \nabla_r \Phi$, where $\vec{A}_\Psi = \vec{0}$. Note that in the relativistic case, the Coulomb gauge can also be used. In this case, the potential $U$ will change, while the quantum potential $Q$ will remain the same, as it is gauge-independent in the Helmholtz decomposition.

From a physical point of view, the presence of the quantum potential $Q$ (i.10) seems interesting for both electrodynamics and gravity problems. The quantum potential is determined by the function $f_1$ (i.10). In the non-relativistic case for a homogeneous initial distribution (1.14), the evolution of the charge or mass density $\rho_s$ is given by expression (1.30), i.e., it is homogeneous in radius at each moment in time. Consequently, in both cases (electromagnetic and gravitational), the quantum potential is zero $Q = 0$. By virtue of (i.20), the quantum pressure will be absent. The equation of motion (i.11) takes the simple form:

$$\frac{d}{dt}\langle \vec{v} \rangle = \left( \frac{\partial}{\partial t} + \langle \vec{v} \rangle \cdot \nabla_r \right) \langle \vec{v} \rangle = -\frac{1}{m} \nabla_r U = \langle \langle \dot{\vec{v}} \rangle \rangle, \qquad (3.5)$$

$$U(r,t) = \frac{mr^2}{2} \begin{cases} -\left[ \dot{b}_e(t) + b_e^2(t) \right], & EM-case, \\ \dot{b}_g(t) + b_g^2(t), & G-case, \end{cases} \qquad \langle \vec{v} \rangle(r,t) = \vec{r} \begin{cases} b_e(t), & EM-case, \\ -b_g(t), & G-case, \end{cases} \qquad (3.6)$$

where the functions $b_e(t)$, $b_g(t)$ are positive, monotonically increasing functions and, according to (1.19), (2.11), are proportional to $\sqrt{1 - P_s^{(em)}(\lambda_s t)^{-1}}$ and $\sqrt{P_s^{(g)}(\lambda_s t)^{-1} - 1}$, respectively. Substituting the quadratic potential $U \sim r^2$ (3.6) into the Vlasov-Moyal approximation (i.23) yields an expression for $m\langle \dot{\vec{v}} \rangle = -\nabla_r U = m\langle \langle \dot{\vec{v}} \rangle \rangle$, taking into account (3.5). In the electromagnetic case, the potential $U$ is repulsive with a stiffness coefficient decreasing in time (3.6), and the system resembles the problem of wave packet spreading in quantum mechanics. In the gravitational case, the potential $U$ is compressive with a stiffness coefficient increasing in time (3.6), and the system resembles the problem of a quantum harmonic oscillator with a variable frequency. Note that in the usual time independent problem of a quantum harmonic oscillator, there is a quantum pressure counteracting the external force $-\nabla_r U$. In this case (of a homogeneous distribution), compression and expansion occur at a certain «equilibrium» rate that does not lead to a back-reaction of the system (in the form of quantum pressure) to the external influence $-\nabla_r U$.

Accounting for relativism fundamentally changes the behavior of both systems. As seen in Fig. 4, left, the density distribution is not homogeneous. The emerging slope (see §1) leads to the presence of a quantum potential $Q \neq 0$ and, consequently, to the presence of a quantum pressure force $-\nabla_r Q$, counteracting the external force $-\nabla_r U$. A similar situation arises in the gravitational case. As seen in Fig. 8, at large times the mass density differs significantly from a constant. As a result, a quantum pressure force arises, counteracting the compressive force. For inhomogeneous initial distributions of type (1.15), quantum pressure is present in both relativistic and non-relativistic cases (see Fig. 4, right, and Fig. 9). In the classical limit $\hbar \to 0$, the quantum pressure effect vanishes.

Note that by formally constructing the «wave function» $\Psi$ for both the electromagnetic and gravitational systems, one can find the Wigner function $W(\vec{r}, \vec{p}, t)$ (i.21). The obtained function $W$ may not coincide with the original function $f_2$, since the potential $U$ depends on the gauge freedom in the Helmholtz decomposition. There are infinitely many functions $f_2$ that, upon averaging, yield the fixed functions $f_1$ and $\langle \vec{v} \rangle$. As mentioned earlier (see Introduction), the Wigner function possesses negative values for all $\Psi$ different from a Gaussian distribution. The area of the phase space region with negative values is of the order of $\sim \hbar$, which is significantly smaller than the area with positive values.

The presence of quantum pressure and the sign-alternating nature of the Wigner function may draw analogies with concepts such as dark energy and dark matter in gravity theory. However, it should be remembered that GR is based on a geometric model that has moved away from the potential theory to which quantum mechanics and the Vlasov chain of equations belong.

**Conclusions**

The exact solutions obtained in this work are useful for modeling beam dynamics in accelerator physics. Simulating charged particle beams with high density requires accounting for the space charge effect. The significant nonlinearity of the system, which requires considering the influence of each particle on every other, leads to a numerical solution of the problem. When using various methods like PP or PIC (PP: Particle-to-Particle, PIC: Particle-In-Cell), questions always arise regarding the choice of the charge density assignment algorithm, the Eulerian grid step, the correctness of solving the equations of motion on the Lagrangian grid, and the subsequent recalculation of electromagnetic field distributions. The existence of exact analytical solutions for a system of many interacting particles allows for a cross-check of the numerical software code and optimization of the algorithm parameters using known test cases.

The system with cylindrical symmetry can correspond to a continuous longitudinal beam, and the system with spherical symmetry can be useful for modeling a bunched beam in an accelerator. Exact solutions to the problems with the Coulomb explosion in plasma physics are also of interest. The gravitational solutions from §2 are a natural mathematical extension of the electromagnetic problem due to the similarity of the interaction potentials and can be useful for modeling astrophysical problems. The connection between the solutions of continuum mechanics in §1 and quantum theory is not coincidental, as quantum systems with a shock wave are also well-known [57-59].

As demonstrated in this work using the obtained exact solutions, the mathematical formalism of Wigner and Vlasov allows for a natural connection between physical systems from various areas of physics within a single representation – through the distribution function.

**Appendix A**

*Proof of Theorem 1*

Let us transform the original equation (1.2). We obtain

$$\frac{d}{dt}\frac{\dot{R}}{\sqrt{1-\dot{R}^2/c^2}} = \frac{\ddot{R}c^2\sqrt{1-\dot{R}^2/c^2}\sqrt{1-\dot{R}^2/c^2} + \dot{R}^2\ddot{R}}{c^2\sqrt{1-\dot{R}^2/c^2}\left(1-\dot{R}^2/c^2\right)} = \frac{\ddot{R}}{\left(1-\dot{R}^2/c^2\right)^{3/2}} = \frac{\vartheta_s^2}{R^2},$$

$$\ddot{R} = \frac{\vartheta_s^2}{R^2}\left(1-\dot{R}^2/c^2\right)^{3/2}. \tag{A.1}$$

Multiply equation (A.1) by $\dot{R}$

$$\dot{R}\ddot{R} = \vartheta_s^2 \frac{\dot{R}}{R^2}\left(1-\dot{R}^2/c^2\right)^{3/2} = -\vartheta_s^2\left(1-\dot{R}^2/c^2\right)^{3/2}\frac{d}{dt}\frac{1}{R},$$

$$\left(1-\dot{R}^2/c^2\right)^{-3/2}\dot{R}\ddot{R}=-\vartheta_s^2\frac{d}{dt}\frac{1}{R}. \tag{A.2}$$

Note that

$$-\frac{c^2}{2n}\frac{d}{dt}\left(1-\dot{R}^2/c^2\right)^n=\left(1-\dot{R}^2/c^2\right)^{n-1}\dot{R}\ddot{R}. \tag{A.3}$$

Expression (A.3) allows us to transform equation (A.2) for $n=-1/2$

$$\frac{d}{dt}\left(1-\dot{R}^2/c^2\right)^{-1/2}=-\bar{\beta}_s^2\frac{d}{dt}\frac{1}{R},$$

$$\left(1-\dot{R}^2/c^2\right)^{-1/2}=C_0-\bar{\beta}_s^2\frac{1}{R}\Rightarrow\frac{\dot{R}}{c}=\pm\sqrt{1-\frac{1}{\left(C_0-\bar{\beta}_s^2\frac{1}{R}\right)^2}}, \tag{A.4}$$

where $C_0$ is a constant. Note that the sign in front of the root in (A.4) must be «+» because the radius $R$ is increasing. From (A.5) the initial condition follows (1.5). Let us find the solution of equation (A.4), we get

$$\frac{\dot{R}}{c}=\frac{R}{C_0 R-\bar{\beta}_s^2}\sqrt{\left(\frac{C_0 R-\bar{\beta}_s^2}{R}\right)^2-1}=\frac{1}{C_0 R-\bar{\beta}_s^2}\sqrt{\left(C_0^2-1\right)R^2-2\bar{\beta}_s^2 C_0 R+\bar{\beta}_s^4},$$

$$\frac{\left(C_0 R-\bar{\beta}_s^2\right)dR}{\sqrt{\left(C_0^2-1\right)R^2-2\bar{\beta}_s^2 C_0 R+\bar{\beta}_s^4}}=cdt,\quad \frac{\left(C_0 R-\bar{\beta}_s^2\right)d\left(C_0 R\right)}{\sqrt{\frac{C_0^2-1}{C_0^2}\left(C_0 R-\frac{\bar{\beta}_s^2 C_0^2}{C_0^2-1}\right)^2-\frac{\bar{\beta}_s^4}{C_0^2-1}}}=C_0 cdt,$$

$$\frac{\left(C_0 R-\bar{\beta}_s^2\right)d\left(C_0 R\right)}{\sqrt{\left(C_0 R-\frac{\bar{\beta}_s^2 C_0^2}{C_0^2-1}\right)^2-\left(\frac{\bar{\beta}_s^2 C_0}{C_0^2-1}\right)^2}}=cdt\sqrt{C_0^2-1}. \tag{A.5}$$

For the convenience of mathematical transformations, let us introduce the notations $b=\frac{\bar{\beta}_s^2 C_0}{C_0^2-1}$, $a=bC_0$ and $y=C_0 R$. Then equation (A.5) takes the form

$$\frac{\left(y-\bar{\beta}_s^2\right)dy}{\sqrt{\left(y-a\right)^2-b^2}}=cdt\sqrt{C_0^2-1}, \tag{A.6}$$

hence

$$\sqrt{\left(y-a\right)^2-b^2}+\left(a-\bar{\beta}_s^2\right)\text{arcch}\frac{y-a}{b}+C_1=ct\sqrt{C_0^2-1}, \tag{A.7}$$

where $C_1$ is a constant. Performing the reverse change of variables in expression (A.7), we obtain

$$\sqrt{C_0^2 R^2 - 2\frac{\overline{\beta}_s^2 C_0^3}{C_0^2 - 1} R + \frac{\overline{\beta}_s^4 C_0^2}{C_0^2 - 1}} - \overline{\beta}_s^2 \left(1 - \frac{C_0^2}{C_0^2 - 1}\right) \operatorname{arcch} C_0 \left(\frac{C_0^2 - 1}{\overline{\beta}_s^2 C_0} R - 1\right) + C_1 = ct\sqrt{C_0^2 - 1}. \quad (A.8)$$

Let us factorize the sub-root expression in (A.8)

$$R^2 - 2\frac{\overline{\beta}_s^2 C_0}{C_0^2 - 1} R + \frac{\overline{\beta}_s^4}{C_0^2 - 1} = \left(R - \frac{\overline{\beta}_s^2}{C_0 - 1}\right)\left(R - \frac{\overline{\beta}_s^2}{C_0 + 1}\right). \quad (A.9)$$

Taking (A.9) into account, solution (A.8) takes the form

$$C_0 \sqrt{\left(R - \frac{\overline{\beta}_s^2}{C_0 - 1}\right)\left(R - \frac{\overline{\beta}_s^2}{C_0 + 1}\right)} + \frac{\overline{\beta}_s^2}{C_0^2 - 1} \operatorname{arcch} C_0 \left(\frac{C_0^2 - 1}{\overline{\beta}_s^2 C_0} R - 1\right) + C_1 = ct\sqrt{C_0^2 - 1}. \quad (A.10)$$

Expression (A.10) implies the initial condition (1.6) and the solution (1.4). Theorem 1 is proved.

*Proof of Corollary 1*
Direct substitution $\dot{R}_0 = 0$ into expressions (1.5) and (1.6) yields expressions (1.8) for $C_0, C_1$. Substituting the values into $C_0, C_1$ in (1.4), we obtain

$$\sqrt{\left(\frac{R}{R_0} - 1\right)\left(\frac{R}{R_0} - \frac{\eta_s^2}{\eta_s^2 + 2}\right)} + \frac{1}{(1 + \eta_s^2)(\eta_s^2 + 2)} \operatorname{arcch}\left[(\eta_s^2 + 2)\frac{R}{R_0} - 1 - \eta_s^2\right] = t\frac{c\eta_s\sqrt{\eta_s^2 + 2}}{R_0(1 + \eta_s^2)}, \quad (A.11)$$

hence

$$\sqrt{(x-1)\left(x - \frac{y}{y+2}\right)} + \frac{\operatorname{arcch}\left[(y+2)x - 1 - y\right]}{(y+2)(1+y)} = \lambda_s(R_0)t, \quad x = \frac{R}{R_0}, \ y = \eta_s^2, \quad (A.12)$$

where $C_0^2 - 1 = \eta_s^4 + 2\eta_s^2 = \eta_s^2(\eta_s^2 + 2)$. Corollary 1 is proved.

*Proof of Corollary 2*
Let us now consider the limit transition

$$\lim_{c \to +\infty} \eta_s^2(R_0) = \lim_{c \to +\infty} \frac{\vartheta_s^2(R_0)}{c^2 R_0} = 0, \quad \lim_{c \to +\infty} \lambda_s(R_0) = \frac{\vartheta_s(R_0)}{R_0^{3/2}} \lim_{c \to +\infty} \frac{\sqrt{2 + \eta_s^2(R_0)}}{1 + \eta_s^2(R_0)} = \sqrt{2}\frac{\vartheta_s(R_0)}{R_0^{3/2}},$$

$$\lim_{c \to +\infty} \mathcal{F}_s(x, \eta_s^2) = \sqrt{x(x-1)} + \frac{1}{2}\operatorname{arcch}(2x - 1). \quad (A.13)$$

Note that $\operatorname{arcch}(x) = \ln\left(x + \sqrt{x^2 - 1}\right)$, and therefore,

$$\frac{1}{2}\operatorname{arcch}(2x-1) = \frac{1}{2}\ln\left(2x-1+\sqrt{(2x-1)^2-1}\right) = \frac{1}{2}\ln\left(x+2\sqrt{x(x-1)}+x-1\right) =$$
$$= \frac{1}{2}\ln\left(\sqrt{x}+\sqrt{x-1}\right)^2 = \ln\left(\sqrt{x}+\sqrt{x-1}\right) = \operatorname{arcch}\sqrt{x}. \tag{A.14}$$

Taking into account (A.13) and (A.14), we establish the validity of (1.11), (1.12). Corollary 2 is proved.

### Proof of Corollary 3

From expression (A.4) it follows that:

$$\beta = \frac{1}{C_0 - \bar{\beta}_s^2/R}\sqrt{\left(C_0 - \bar{\beta}_s^2/R\right)^2 - 1}, \tag{A.15}$$

where according to (1.8) $C_0 = 1 + \bar{\beta}_s^2(R_0)/R_0$, consequently,

$$C_0 - \bar{\beta}_s^2/R = 1 + \eta_s^2(R_0)\kappa_s(t, R_0), \tag{A.16}$$

$$\kappa_s(t, R_0) \stackrel{\text{def}}{=} \frac{R - R_0}{R} = \frac{R/R_0 - 1}{R/R_0} = \frac{\mathcal{P}_s\left[\lambda_s(R_0)t, \eta_s^2(R_0)\right] - 1}{\mathcal{P}_s\left[\lambda_s(R_0)t, \eta_s^2(R_0)\right]}. \tag{A.17}$$

Substituting expression (A.16) into (A.15), we obtain the validity of (1.16)-(1.17). Let us perform the limit transition for (1.16) as $c \to +\infty$, we obtain

$$\mathrm{v}_s = \lim_{c \to \infty} c\beta_s(t, R_0) = \lim_{c \to \infty} c\eta_s(R_0)\sqrt{2\kappa_s(t, R_0)} = R_0\lambda_s(R_0)\sqrt{\frac{\mathcal{P}_s\left[\lambda_s(R_0)t\right] - 1}{\mathcal{P}_s\left[\lambda_s(R_0)t\right]}}, \tag{A.18}$$

where the relations (1.13) and $\lim_{c \to \infty}\eta_s = 0$ are taken into account. Let us derive the expressions for the asymptotics (1.17) and (1.20). Since the size of the spherical layer increases monotonically with time ($R(t) \geq R_0$), then $\mathcal{P}_s \to +\infty$. Consequently,

$$\lim_{t \to +\infty}\kappa_s(t, R_0) = \lim_{t \to +\infty}\left\{1 - \frac{1}{\mathcal{P}_s\left[\lambda_s(R_0)t, \eta_s^2(R_0)\right]}\right\} = 1. \tag{A.19}$$

Substituting (A.19) into (1.16) for the limit transition, we obtain (1.17). Similarly to expression (A.19) the limit is taken for (1.19), i.e. $P_s \to +\infty$ as $t \to +\infty$. As a result, expression (1.19) transforms into (1.20). Corollary 3 is proved.

### Proof of Theorem 2

According to expression (1.1), the charge $\Delta Q_s$, contained between the layers $R_1(t) = r\mathcal{P}_s\left[\lambda_s(r)t, \eta_s^2(r)\right]$ and $R_2(t) = (r+\Delta r)\mathcal{P}_s\left[\lambda_s(r+\Delta r)t, \eta_s^2(r+\Delta r)\right]$ over the time interval $t < t_c$ remains unchanged, consequently

$$\Delta Q_s = 4\pi\int_r^{r+\Delta r}\rho_0(x)x^2 dx = 4\pi\int_{R_1(t)}^{R_2(t)}\rho_s(x,t)x^2 dx = 4\pi\rho_0(r)r^2\Delta r + O(\Delta r^2), \tag{A.20}$$

$$\Delta R(t) = R_2(t) - R_1(t) = (r + \Delta r) \mathcal{P}_s \left[\lambda_s(r+\Delta r)t, \eta_s^2(r+\Delta r)\right] - r\mathcal{P}_s\left[\lambda_s(r)t, \eta_s^2(r)\right] =$$

$$= \Delta r \frac{\partial}{\partial x}\left\{x\mathcal{P}_s\left[\lambda_s(x)t, \eta_s^2(x)\right]\right\}\bigg|_{x=r} + O(\Delta r^2).$$

The charge density $\rho_s$ is determined by the relation $\Delta Q_s / \Delta R$, therefore, taking (A.20) into account, we obtain

$$\frac{dQ_s}{dR} = \lim_{\Delta R \to 0} \frac{\Delta Q_s}{\Delta R} = \frac{4\pi \rho_0(r) r^2}{\mathcal{P}_s\left[\lambda_s(r)t, \eta_s^2(r)\right] + r\frac{\partial}{\partial r}\mathcal{P}_s\left[\lambda_s(r)t, \eta_s^2(r)\right]} = 4\pi \rho_s(R,t) R^2, \quad (A.21)$$

hence

$$\rho_s(R,t) = \frac{\rho_0(r)}{\mathcal{P}_s^2\left[\lambda_s(r)t, \eta_s^2(r)\right]\left\{\mathcal{P}_s\left[\lambda_s(r)t, \eta_s^2(r)\right] + r\frac{\partial}{\partial r}\mathcal{P}_s\left[\lambda_s(r)t, \eta_s^2(r)\right]\right\}}, \quad (A.22)$$

where it is taken into account that (1.10) $R(r,t) = r\mathcal{P}_s\left[\lambda_s(r)t, \eta_s^2(r)\right]$. Let us compute the derivative $\partial_r \mathcal{P}_s$, entering expression (A.22). Differentiate expression (1.7), we obtain

$$\lambda_s'(r)t = \frac{rR_r'(t,r) - R(t,r)}{r^2}\frac{\partial}{\partial x}\mathcal{F}_s(x,y) + 2\eta_s \eta_s'(r)\frac{\partial}{\partial y}\mathcal{F}_s(x,y),$$

$$R_r'(t,r) = r\frac{\lambda_s'(r)t - 2\eta_s \eta_s'(r)\partial_y \mathcal{F}_s(x,y)}{\partial_x \mathcal{F}_s(x,y)} + \frac{R(t,r)}{r}. \quad (A.23)$$

The expression in the denominator of (A.22), taking (A.23) into account, takes the form

$$\frac{\partial}{\partial r}R\left[t, \eta_s^2(r)\right] = \mathcal{P}_s\left[\lambda_s(r)t, \eta_s^2(r)\right] + r\frac{\partial}{\partial r}\mathcal{P}_s\left[\lambda_s(r)t, \eta_s^2(r)\right] =$$

$$= r\frac{\lambda_s'(r)t - 2\eta_s \eta_s'(r)\partial_{\eta_s^2}\mathcal{F}_s(\mathcal{P}_s, \eta_s^2)}{\partial_{\mathcal{P}_s}\mathcal{F}_s(\mathcal{P}_s, \eta_s^2)} + \mathcal{P}_s(\lambda_s t, \eta_s^2). \quad (A.24)$$

Substituting (A.24) into (A.22) proves the validity of expression (1.23). Since the function $\mathcal{F}_s$ is known explicitly (1.9), let us compute the derivatives $\partial_{\eta_s^2}\mathcal{F}_s$ and $\partial_{\mathcal{P}_s}\mathcal{F}_s$, entering expression (1.23), we obtain:

$$\partial_x \mathcal{F}_s(x,y) = \frac{2x - 1 - \frac{y}{y+2}}{2\sqrt{(x-1)\left(x - \frac{y}{y+2}\right)}} + \frac{1}{(y+1)\sqrt{\left[(y+2)x - y - 1\right]^2 - 1}} = \quad (A.25)$$

$$= \left[(x-1)\left(x - \frac{y}{y+2}\right)\right]^{-1/2}\left[\frac{1}{2}\left(2x - 1 - \frac{y}{y+2}\right) + \frac{1}{(y+1)(y+2)}\right] =$$

$$= \left[(x-1)\left(x - \frac{y}{y+2}\right)\right]^{-1/2}\frac{\left[(2x-1)(y+2) - y\right](y+1) + 2}{2(y+1)(y+2)}.$$

Performing a regrouping of terms in the numerator of expression (A.25), we obtain expression (1.26). Let us find the derivative $\partial_y \mathcal{F}_s(x,y)$. From direct computations it follows

$$\partial_y \mathcal{F}_s(x,y) = \frac{\sqrt{x-1}}{(y+2)^2 \sqrt{x-\frac{y}{y+2}}} \left(\frac{1}{y+1} - 1\right) - \frac{(2y+3)\operatorname{arcch}\left[(y+2)x - y - 1\right]}{(y+1)^2 (y+2)^2}. \tag{A.26}$$

From expression (A.26), the validity of (1.27) follows directly. For derivatives (1.24)-(1.25), we obtain

$$2\vartheta_s \vartheta_s' = \frac{qQ_s'}{4\pi\varepsilon_0 m} = \frac{q}{\varepsilon_0 m} r^2 \rho_0(r) \Rightarrow \overline{\beta}_s' = \frac{\vartheta_s'}{c} = \frac{qr^2 \rho_0(r)}{2mc\varepsilon_0 \vartheta_s(r)} = \frac{qr^2 \rho_0(r)}{2mc^2 \varepsilon_0 \overline{\beta}_s(r)}, \tag{A.27}$$

$$\eta_s' = \frac{2\overline{\beta}_s' r - \overline{\beta}_s}{2r^{3/2}} = \frac{qr^3 \rho_0(r) - mc^2 \varepsilon_0 \overline{\beta}_s^2(r)}{2mc^2 \varepsilon_0 r^{3/2} \overline{\beta}_s(r)} = \frac{qr^2 \rho_0(r) - mc^2 \varepsilon_0 \eta_s^2(r)}{2mc^2 \varepsilon_0 r \eta_s(r)}, \tag{A.28}$$

hence

$$2\eta_s \eta_s' = \frac{qr^2 \rho_0(r) - mc^2 \varepsilon_0 \eta_s^2(r)}{mc^2 \varepsilon_0 r} = \frac{qr \rho_0(r)}{\varepsilon_0 mc^2} - \frac{\eta_s^2(r)}{r}. \tag{A.29}$$

Expression (A.29) coincides with (1.24). From expression (1.8) we find $\lambda_s'(r)$

$$\lambda_s'(r) = c \frac{2r(1+\eta_s^2)^2 \eta_s' - \eta_s(2+\eta_s^2)(1+\eta_s^2 + r 2\eta_s \eta_s')}{r^2 (1+\eta_s^2)^2 \sqrt{2+\eta_s^2}} = c \frac{2r\eta_s' - \eta_s(2+\eta_s^2)(1+\eta_s^2)}{r^2 (1+\eta_s^2)^2 \sqrt{2+\eta_s^2}}. \tag{A.30}$$

As a result, substituting (A.28) into (A.30), we obtain expression (1.25). Theorem 2 is proved.

*Proof of Corollary 4*
The limit transitions for $\eta_s(r)$, $\lambda_s(r)$ and $\mathcal{P}_s$ were considered earlier (A.13), (1.13), so it remains to consider the limits only for $\lambda_s'$, $\eta_s'$, $\partial_{\eta_s^2}\mathcal{F}_s(\mathcal{P}_s, \eta_s^2)$ and $\partial_{\mathcal{P}_s}\mathcal{F}_s(\mathcal{P}_s, \eta_s^2)$. From expression (1.25) it follows:

$$\lim_{c \to +\infty} \lambda_s'(r) = \frac{q\rho_0(r)}{\varepsilon_0 m\sqrt{2}} \lim_{c \to +\infty} \frac{1}{c\eta_s(r)} - \frac{3}{r^2 \sqrt{2}} \lim_{c \to +\infty} c\eta_s(r) = \frac{q\rho_0(r)}{\varepsilon_0 m\sqrt{2}} \frac{\sqrt{r}}{\vartheta_s(r)} - \frac{3}{r^2 \sqrt{2}} \frac{\vartheta_s(r)}{\sqrt{r}} =$$
$$= \frac{1}{r}\left[\frac{q\rho_0(r)}{m\varepsilon_0 \lambda_s(r)} - \frac{3}{2}\lambda_s(r)\right]. \tag{A.31}$$

Let us compute the limit of the expression $\eta_s(r)\eta_s'(r)\partial_{\eta_s^2}\mathcal{F}_s(\mathcal{P}_s, \eta_s^2)$, we obtain:

$$\lim_{c \to +\infty} \partial_{\eta_s^2}\mathcal{F}_s(\mathcal{P}_s, \eta_s^2) = \lim_{y \to 0} \partial_y \mathcal{F}_s(x,y) = -\frac{3}{4}\operatorname{arcch}(2x-1), \tag{A.32}$$

hence

$$\lim_{c\to+\infty}\left[2\eta_s(r)\eta_s'(r)\partial_{\eta_s^2}\mathcal{F}_s(\mathcal{P}_s,\eta_s^2)\right]=-\frac{3}{4}\operatorname{arcch}(2x-1)\left[\lim_{c\to+\infty}\frac{qr\rho_0(r)}{\varepsilon_0 mc^2}-\lim_{c\to+\infty}\frac{\vartheta_s^2(r)}{r^2c^2}\right]=0. \quad (A.33)$$

The limit of $\partial_{\mathcal{P}_s}\mathcal{F}_s(\mathcal{P}_s,\eta_s^2)$ in accordance with (1.27) has the form

$$\lim_{c\to+\infty}\partial_{\mathcal{P}_s}\mathcal{F}_s(\mathcal{P}_s,\eta_s^2)=\lim_{y\to 0}\partial_x\mathcal{F}_s(x,y)=\left[(x-1)x\right]^{-1/2}\;x=\sqrt{\frac{x}{x-1}}=\sqrt{\frac{P_s[\lambda_s(r)t]}{P_s[\lambda_s(r)t]-1}}. \quad (A.34)$$

Substituting expressions (A.34), (A.33), and (A.31) into the limit transition for the density $\lim_{c\to+\infty}\rho_s(R,t)$, we obtain the validity of expression (1.29). Corollary 4 is proved.

*Proof of Lemma 1*

Note that the charge motion satisfies the continuity equation $\partial_t\rho_s+\operatorname{div}_r[\rho_s\langle\vec{v}\rangle]=0$. Integrating it over the volume of a sphere of radius $r$, we obtain

$$4\pi\frac{\partial}{\partial t}\int_0^r x^2\rho_s(x,t)dx+\int_{S_r}\rho_s\langle\vec{v}\rangle d\vec{s}_r=0, \quad (A.35)$$

$$4\pi\frac{\partial}{\partial t}Q_s(r,t)+\rho_s(r,t)\langle v\rangle(r,t)4\pi r^2=0, \quad (A.36)$$

where $S_r$ is surface of the sphere with radius $r$. When transitioning from (A.35) to (A.36), the symmetry of the charge density distribution $\rho_s$, is taken into account, i.e., $\langle\vec{v}\rangle\parallel d\vec{s}_r$ and $J=\rho_s\langle v\rangle$ is constant on the spherical surface. Since $\partial_r Q_s(r,t)=4\pi r^2\rho_s(r,t)$, equation (A.36) transforms into equation (1.31). The obtained equation $dQ_s/dt=0$ admits (see Remark 4 and (1.28)) a solution in the form of characteristics (1.10). Lemma 1 is proved.

*Proof of Theorem 3*

Let us transform the equation of motion (1.33). We obtain

$$\ddot{R}=\frac{\vartheta_c^2}{R}(1-\dot{R}^2/c^2)^{3/2}\Rightarrow\dot{R}\ddot{R}=\vartheta_c^2\frac{\dot{R}}{R}(1-\dot{R}^2/c^2)^{3/2}=\vartheta_c^2(1-\dot{R}^2/c^2)^{3/2}\frac{d}{dt}\ln R,$$
$$(1-\dot{R}^2/c^2)^{-3/2}\dot{R}\ddot{R}=\vartheta_c^2\frac{d}{dt}\ln R\Rightarrow\frac{d}{dt}(1-\dot{R}^2/c^2)^{-1/2}=\bar{\beta}_c^2\frac{d}{dt}\ln R, \quad (A.37)$$

where expression (A.3) is taken into account. Integrating equation (A.37), we obtain

$$(1-\dot{R}^2/c^2)^{-1/2}=C_0+\bar{\beta}_c^2\ln R\Rightarrow\frac{\dot{R}}{c}=\sqrt{1-\frac{1}{(C_0+\bar{\beta}_c^2\ln R)^2}}, \quad (A.38)$$

where the constant $C_0$ is determined from the initial conditions $R(0)=R_0$ and $\dot{R}(0)=0$:

$$C_0+\bar{\beta}_c^2\ln R_0=1\Rightarrow C_0=1-\bar{\beta}_c^2\ln R_0. \quad (A.39)$$

Substituting (A.39) into equation (A.38), we obtain

$$\left[1-\left(1+\bar{\beta}_c^2 \ln \frac{R}{R_0}\right)^{-2}\right]^{-1/2} d\frac{R}{R_0} = \frac{cdt}{R_0} \Rightarrow \frac{1+\bar{\beta}_c^2 \ln x}{\sqrt{(1+\bar{\beta}_c^2 \ln x)^2 - 1}} dx = \frac{cdt}{R_0}. \tag{A.40}$$

Let us introduce the change of variables $\ln x = z^2$. As a result, (A.40) takes the form

$$\frac{1+\bar{\beta}_c^2 z^2}{\sqrt{2+\bar{\beta}_c^2 z^2}} e^{z^2} dz = \frac{\bar{\beta}_c cdt}{2R_0} = \frac{\vartheta_c dt}{2R_0} \Rightarrow \sqrt{2}\int_0^{\sqrt{\ln(R/R_0)}} \frac{1+\bar{\beta}_c^2 z^2}{\sqrt{2+\bar{\beta}_c^2 z^2}} e^{z^2} dz = \frac{\vartheta_c t}{R_0\sqrt{2}}, \tag{A.41}$$

where, by virtue of the initial condition $R(0) = R_0$, the constant of integration is zero. Expression (A.41) proves the validity of (1.34)-(1.35). Let us perform the limit transition $c \to +\infty$ for the function $\mathcal{F}_c(x,y)$. Since $\lim_{c \to +\infty} \bar{\beta}_c = 0$, then $\mathcal{F}_c(x,y) \to F_c(x)$. Theorem 3 is proved.

*Proof of Theorem 4*

By analogy with the proof of Theorem 2, we obtain

$$\Delta Q_c = 2\pi \rho_0(r) r \Delta r + O(\Delta r^2), \quad \Delta R(t) = \Delta r \frac{\partial}{\partial x}\{x\mathcal{P}_c[\lambda_c(x)t, \bar{\beta}_c^2(x)]\}\bigg|_{x=r} + O(\Delta r^2), \tag{A.42}$$

hence

$$\frac{dQ_c}{dR} = \frac{2\pi \rho_0(r) r}{\mathcal{P}_c[\lambda_c(r)t, \bar{\beta}_c^2(r)] + r\frac{\partial}{\partial r}\mathcal{P}_c[\lambda_c(r)t, \bar{\beta}_c^2(r)]} = 2\pi \rho_c(R,t) R, \tag{A.43}$$

consequently,

$$\rho_c(R,t) = \frac{\rho_0(r)}{\mathcal{P}_c^2[\lambda_c(r)t, \bar{\beta}_c^2(r)]\left\{\mathcal{P}_c[\lambda_c(r)t, \bar{\beta}_c^2(r)] + r\frac{\partial}{\partial r}\mathcal{P}_c[\lambda_c(r)t, \bar{\beta}_c^2(r)]\right\}}. \tag{A.44}$$

By analogy with the proof of Theorem 2, for the derivative $\partial_r \mathcal{P}_c$ the expression is valid

$$\frac{\partial}{\partial r}\mathcal{P}_c[\lambda_c(r)t, \bar{\beta}_c^2(r)] = \frac{\lambda_c'(r)t - 2\bar{\beta}_c\bar{\beta}_c'(r)\partial_{\bar{\beta}_c^2}\mathcal{F}_c(\mathcal{P}_s, \bar{\beta}_c^2)}{\partial_{\mathcal{P}_c}\mathcal{F}_c(\mathcal{P}_c, \bar{\beta}_c^2)}. \tag{A.45}$$

Substituting (A.45) into (A.44) gives expression (1.37). The function is known in implicit form (1.35), so only the derivative $\partial_{\mathcal{P}_c}\mathcal{F}_c$ has an explicit form, while the derivative $\partial_y \mathcal{F}_c$ has an implicit form. Indeed, direct differentiation of expression (1.35) leads to the results

$$\frac{\partial}{\partial x}\mathcal{F}_c(x,y) = \frac{\sqrt{2}}{2x\sqrt{\ln x}}\frac{1+y\ln x}{\sqrt{2+y\ln x}}e^{\ln x} = \frac{1+y\ln x}{\sqrt{2\ln x(2+y\ln x)}}, \tag{A.46}$$

$$\frac{\partial}{\partial y}\mathcal{F}_c(x,y) = \sqrt{2}\int_0^{\sqrt{\ln x}}\frac{\partial}{\partial y}\frac{1+yz^2}{\sqrt{2+yz^2}}e^{z^2}dz = \frac{1}{\sqrt{2}}\int_0^{\sqrt{\ln x}}\frac{3+yz^2}{(2+yz^2)^{3/2}}z^2 e^{z^2}dz.$$

Expressions (A.46) coincide with (1.40). Let us compute the derivatives $\bar{\beta}'_c$ and $\lambda'_c$, we obtain

$$2\bar{\beta}_c\bar{\beta}'_c = 2\frac{\vartheta_c\vartheta'_c}{c^2} = \frac{2r\rho_0}{c^2\ell\sqrt{Q_c}}\sqrt{\frac{q\pi}{2\varepsilon_0 m}\frac{qQ_c}{2\pi\varepsilon_0 m}} = \frac{qr\rho_0}{\varepsilon_0\ell mc^2}, \tag{A.47}$$

$$\lambda'_c(r) = \frac{1}{\sqrt{2}}\frac{r\vartheta'_c-\vartheta_c}{r^2} = \frac{1}{\sqrt{2}}\left(\frac{Q'_c}{2r\sqrt{Q_c}}\sqrt{\frac{q}{2\pi\varepsilon_0 m\ell}} - \frac{1}{r^2}\sqrt{\frac{qQ_c}{2\pi\varepsilon_0 m\ell}}\right) = \frac{1}{r^2}\sqrt{\frac{qQ_c}{4\pi\varepsilon_0 m\ell}}\left(\frac{\rho_0\pi r^2}{Q_c}-1\right),$$

where, by virtue of (1.33), it is taken into account that $Q'_c(r) = 2\pi r\rho_0(r)$. Theorem 4 is proved.

*Proof of Corollary 5*

Expression (1.41) is obtained by substituting $\partial_x\mathcal{F}_c(x,y)$ (1.40) into (1.28). Let us take into account that $\mathcal{P}_c \to +\infty$ as $t \to +\infty$, then the limit transition for (1.41) takes the form

$$\beta_c^\infty(R_0) = \bar{\beta}_c(R_0)\lim_{\mathcal{P}_c\to+\infty}\frac{\sqrt{2+\bar{\beta}_c^2(R_0)\ln\mathcal{P}_c}}{\bar{\beta}_c^2(R_0)\sqrt{\ln\mathcal{P}_c}} = \bar{\beta}_c(R_0)\lim_{\mathcal{P}_c\to+\infty}\frac{\sqrt{\bar{\beta}_c^2(R_0)}}{\bar{\beta}_c^2(R_0)} = 1, \tag{A.48}$$

which coincides with (1.42). Let us perform the limit transition for expression (1.37). We obtain

$$\rho_c[R(t,r),t] = \frac{\rho_0}{\mathcal{P}_c\left[\mathcal{P}_c + r\dfrac{t\lambda'_c}{\partial_{\mathcal{P}_c}\mathcal{F}_c(\mathcal{P}_c)}\right]} = \frac{\rho_0}{\mathcal{P}_c\left[\mathcal{P}_c + 2\lambda_c t\sqrt{\ln\mathcal{P}_c}\left(\dfrac{\rho_0\pi r^2}{Q_c}-1\right)\right]}, \tag{A.49}$$

where it is taken into account that $\partial_x F_c(x) = 1/2\sqrt{\ln x}$. Substituting $Q_c/\pi r^2 = 4\varepsilon_0 m\ell\lambda_c^2/q$ (1.35) into expression (A.49), it transforms into (1.43). Let us find the expression for $v$ and $v_\infty$, we obtain

$$v_c(t,R_0) = \vartheta_c(R_0)\sqrt{2\ln\mathcal{P}_c[\lambda_c(R_0)t]} = 2R_0\lambda_c(R_0)\sqrt{\ln\mathcal{P}_c[\lambda_c(R_0)t]}, \tag{A.50}$$

$$v_c^\infty(R_0) = 2R_0\lambda_c(R_0)\lim_{t\to+\infty}\sqrt{\ln\mathcal{P}_c[\lambda_c(R_0)t]} = +\infty,$$

where it is taken into account that $\mathcal{P}_c \to +\infty$ as $t \to +\infty$. Corollary 5 is proved.

**Appendix B**
*Proof of Theorem 5*

According to the proof of Theorem 1 and the initial conditions (1.2), equation (2.1) can be represented in the form

$$\frac{\dot{R}}{c} = -\sqrt{1 - \left[1 + \eta_s^2\left(\frac{R_0}{R} - 1\right)\right]^{-2}}, \tag{B.1}$$

where the minus sign in front of the root (B.1) indicates the decrease of the sphere's radius $R$. Let us transform equation (B.1) by introducing the notation $x = R/R_0$, we obtain

$$\frac{x(1-\eta_s^2)+\eta_s^2}{\sqrt{x^2(\eta_s^2-2)+2(1-\eta_s^2)x+\eta_s^2}} dx = -\frac{\eta_s c}{R_0} dt. \tag{B.2}$$

Let us perform the integration of equation (B.2), first making the following transformations

$$x^2(\eta_s^2-2)+2(1-\eta_s^2)x+\eta_s^2 = \frac{1-z^2}{2-\eta_s^2}, \qquad z \overset{\text{def}}{=} (2-\eta_s^2)\left(x + \frac{\eta_s^2-1}{2-\eta_s^2}\right), \tag{B.3}$$

$$x(1-\eta_s^2)+\eta_s^2 = \frac{1-\eta_s^2}{2-\eta_s^2}\left(z + \frac{1}{1-\eta_s^2}\right).$$

Substituting (B.3) into equation (B.2), we obtain

$$\frac{(1-\eta_s^2)\sqrt{2-\eta_s^2}}{(2-\eta_s^2)^2}\left(\frac{z}{\sqrt{1-z^2}} + \frac{1}{1-\eta_s^2}\frac{1}{\sqrt{1-z^2}}\right) dz = -\frac{\eta_s c}{R_0} dt, \tag{B.4}$$

which after integration takes the form

$$\frac{1-\eta_s^2}{(2-\eta_s^2)^{3/2}}\left(\sqrt{1-z^2} + \frac{\arccos z}{1-\eta_s^2}\right) = \frac{\eta_s c}{R_0} t, \tag{B.5}$$

where the initial condition at $t = 0$, $z = 1$ is taken into account. Producing the reverse change of variables (B.3), we obtain the characteristic equation

$$\sqrt{(1-x)\left(x + \frac{\eta_s^2}{2-\eta_s^2}\right)} + \frac{\arccos\left[(2-\eta_s^2)x + \eta_s^2 - 1\right]}{(1-\eta_s^2)(2-\eta_s^2)} = \frac{\sqrt{2-\eta_s^2}}{1-\eta_s^2}\frac{\eta_s c}{R_0} t, \tag{B.6}$$

where $1-z^2 = -(2-\eta_s^2)^2(x-1)\left(x - \frac{\eta_s^2}{\eta_s^2-2}\right)$ is taken into account. Expression (B.6) coincides with (2.3)-(2.5). Let us perform the limit transition for expression (B.6) as $c \to +\infty$, we obtain

$$\lim_{y \to 0} \mathcal{F}_s(x,y) = \sqrt{(1-x)x} + \frac{1}{2}\arccos(2x-1). \tag{B.7}$$

Taking into account that $\arccos\sqrt{x} = \frac{1}{2}\arccos(2x-1)$, expression (B.7) transforms into (2.6) for the function $F_s(x)$. Theorem 5 is proved.

***Proof of Theorem 6***

Note that when deriving formula (1.23), only the general form of the characteristic equation (1.7) and the expression for the total charge (1.1) are used. Since the characteristic equations (1.7), (2.3) and the expressions for the total charges (1.1) and masses (2.2) are similar, for the problem with gravitational interaction the density function will have the form (1.23). The difference will only be in the derivatives $\eta'_s(r)$, $\lambda'_s(r)$, $\partial_x \mathcal{F}_s(x,y)$ and $\partial_y \mathcal{F}_s(x,y)$. Performing direct computations, we obtain

$$2\eta_s(r)\eta'_s(r) = 2\frac{\bar{\beta}_s}{\sqrt{r}}\frac{2r\bar{\beta}'_s - \bar{\beta}_s}{2r^{3/2}} = \frac{4\pi r \bar{\beta}_s \rho_0 \sqrt{G}}{c\sqrt{M_s}} - \frac{\bar{\beta}_s^2}{r^2} = \frac{4\pi G r \rho_0(r)}{c^2} - \frac{\eta_s^2(r)}{r}, \qquad (B.8)$$

which proves the validity of (2.7). Using (B.8), let us differentiate (2.5)

$$\lambda'_s(r) = c\frac{2r\eta_s\eta'_s - \eta_s^2(2-\eta_s^2)(1-\eta_s^2)}{r^2\eta_s(1-\eta_s^2)^2\sqrt{2-\eta_s^2}} = \frac{4\pi G r^2 \rho_0 - c^2\eta_s^2 - c^2\eta_s^2(2-\eta_s^2)(1-\eta_s^2)}{cr^2\eta_s(1-\eta_s^2)^2\sqrt{2-\eta_s^2}}, \qquad (B.9)$$

which transforms into (2.8). Let us compute the derivatives of the function $\mathcal{F}_s$, we obtain

$$\partial_x \mathcal{F}_s(x,y) = \left[(1-x)\left(x+\frac{y}{2-y}\right)\right]^{-1/2}\left[\frac{-1}{(1-y)|2-y|} + \frac{1}{2}\left(1-2x-\frac{y}{2-y}\right)\right], \qquad (B.10)$$

$$\partial_y \mathcal{F}_s(x,y) = \sqrt{1-x}\left(x+\frac{y}{2-y}\right)^{-1/2}\left[\frac{1}{(2-y)^2} - \frac{1}{(2-y)|2-y|(1-y)}\right] + \\ + (3-2y)\frac{\arccos[(2-y)x-1+y]}{(1-y)^2(2-y)^2}. \qquad (B.11)$$

The argument $y$ in expressions (B.10)-(B.11) corresponds to $\eta_s^2(r) = \frac{GM_s(r)}{c^2 r}$, which in the quasi-relativistic case is less than one. Consequently, the absolute value sign in expressions (B.10) and (B.11) can be removed. As a result, expressions (B.10)-(B.11) transform into expressions (2.9) and (2.10), respectively. Performing the limit transition $c \to +\infty$ in formulas (2.8)-(2.9), we obtain:

$$\lim_{c \to +\infty} \lambda'_s(r) = \frac{4\pi G r^2 \rho_0 - \frac{\vartheta_s^2}{r}3}{r^2\sqrt{\frac{2GM_s}{r}}} = \frac{1}{r}\left[\frac{4\pi G \rho_0(r)}{\lambda_s(r)} - \frac{3}{2}\lambda_s(r)\right], \qquad (B.12)$$

$$\lim_{c \to +\infty} \partial_{\mathcal{P}_s} \mathcal{F}_s(\mathcal{P}_s, \eta_s^2) = \lim_{y \to 0} \partial_x \mathcal{F}_s(x,y) = -\sqrt{\frac{x}{1-x}}. \qquad (B.13)$$

Substituting expressions (B.12) and (B.13) into (A.22), we obtain expression (2.11). Theorem 6 is proved.

*Proof of Theorem 7*

By analogy with the proof of Theorem 3 for equation (2.12), we obtain

$$\frac{\dot{R}}{c} = -\sqrt{1 - \left(C_0 - \bar{\beta}_c^2 \ln R\right)^{-2}}, \quad C_0 = 1 + \bar{\beta}_c^2 \ln R_0, \tag{B.14}$$

hence

$$\frac{1 - \bar{\beta}_c^2 \ln x}{\sqrt{\left(1 - \bar{\beta}_c^2 \ln x\right)^2 - 1}} dx = -\frac{cdt}{R_0}, \tag{B.15}$$

where the minus sign in front of the root (B.14) indicates the direction of velocity towards the center of the sphere. Due to gravitational compression, $x \leq 1$. Consequently, we can introduce the change of variables $\ln x = -z^2$. As a result, (B.15) takes the form

$$\frac{1 + \bar{\beta}_c^2 z^2}{\sqrt{2 + \bar{\beta}_c^2 z^2}} e^{-z^2} dz = \frac{\bar{\beta}_c cdt}{2R_0} = \frac{\vartheta_c dt}{2R_0} \Rightarrow \sqrt{2} \int_0^{\sqrt{\ln(R_0/R)}} \frac{1 + \bar{\beta}_c^2 z^2}{\sqrt{2 + \bar{\beta}_c^2 z^2}} e^{-z^2} dz = \frac{\vartheta_c t}{R_0 \sqrt{2}}, \tag{B.16}$$

where, by virtue of the initial condition $R(0) = R_0$, the constant of integration is zero. Expression (B.16) proves the validity of (2.13). The limit transition $c \to +\infty$ for the function $\mathcal{F}_c(x, y)$ gives $F_c(x) = \int_0^{\sqrt{\ln(1/x)}} e^{-z^2} dz$. Let us compute the derivatives entering the expression for the mass density function. Let us take into account that $\bar{\beta}_c^2 = 2GM_c/\ell c^2$ and $\vartheta_c^2 = c^2 \bar{\beta}_c^2$, we get

$$2\bar{\beta}_c \bar{\beta}_c' = \frac{4\pi G}{\ell c^2} r \rho_0, \quad 2\vartheta_c \lambda_c' = \frac{c^2}{\sqrt{2}} \frac{r \bar{\beta}_c^{2\prime} - 2\bar{\beta}_c^2}{r^2} = \frac{4G}{r^2 \ell \sqrt{2}} \left(\pi r^2 \rho_0 - M_c\right), \tag{B.17}$$

$$\lambda_c'(r) = \frac{\sqrt{G}}{r^2 \sqrt{\ell M_c}} \left(\pi r^2 \rho_0 - M_c\right) = \frac{1}{r^2} \sqrt{\frac{G}{\ell} M_c} \left(\frac{\pi r^2 \rho_0}{M_c} - 1\right). \tag{B.18}$$

The derivatives of the function $\mathcal{F}_c$ are computed by analogy with (A.46)

$$\partial_x \mathcal{F}_c = -\frac{1 + y \ln(1/x)}{\sqrt{2 + y \ln(1/x)}} \frac{xe^{-\ln(1/x)} \sqrt{2}}{2x^2 \sqrt{\ln(1/x)}} = -\frac{1 - y \ln x}{\sqrt{2 - y \ln x}} \frac{1}{\sqrt{-2 \ln x}} = \frac{y \ln x - 1}{\sqrt{2 \ln x (y \ln x - 2)}}. \tag{B.19}$$

The derivative $\partial_y \mathcal{F}_c$ will be similar to the analogous derivative from (A.46), differing only in the sign in the exponent's power. Expression (2.17) follows directly from substituting (B.18) and (B.19) for $y \to 0$ into expression (1.37). A similar procedure was done earlier in the proof of Corollary 5. Theorem 7 is proved.

*Gauge condition*

Let us differentiate the Hamilton-Jacobi equation and apply the $\Psi$-gauge. We obtain

$$\frac{1}{c^2}\frac{\partial^2 \varphi}{\partial t^2} - \frac{\alpha_2}{2\alpha_1}\operatorname{div}_r \vec{A}_\Psi = \frac{1}{4\alpha_1 c^2}\frac{\partial}{\partial t}\left|\langle \vec{v}\rangle\right|^2. \qquad (B.20)$$

From the Helmholtz decomposition it follows that $\alpha_2 \operatorname{div}_r \vec{A}_\Psi = \operatorname{div}_r \langle \vec{v}\rangle + 2\alpha_1 \Delta_r \varphi$, hence equation (B.20) takes the form:

$$\frac{1}{c^2}\frac{\partial^2 \varphi}{\partial t^2} - \Delta_r \varphi = \frac{1}{4\alpha_1 c^2}\frac{\partial}{\partial t}\left|\langle \vec{v}\rangle\right|^2 + \frac{1}{2\alpha_1}\operatorname{div}_r \langle \vec{v}\rangle. \qquad (B.21)$$